%% file: main.tex
\documentclass[fleqn,usenatbib]{mnras}

\usepackage{newtxtext,newtxmath,times}

\usepackage[T1]{fontenc}
\usepackage{ae,aecompl}
\usepackage{cases}

\DeclareRobustCommand{\VAN}[3]{#2}
\let\VANthebibliography\thebibliography
\def\thebibliography{\DeclareRobustCommand{\VAN}[3]{##3}\VANthebibliography}

\usepackage{graphicx}	
\usepackage{amsmath}	

\newcommand{\ramses}{\textsc{ramses}}
\newcommand{\exahype}{\textsc{ExaHyPE}}

\title[Self-similar accretion in different gravity models]
{Spherical accretion of collisional gas in modified gravity I: self-similar solutions and a new cosmological hydrodynamical code}

\author[H. Zhang et al.]{
Han Zhang,$^{1}$\thanks{E-mail: han.zhang3@durham.ac.uk}
Tobias Weinzierl,$^{2,3}$
Holger Schulz$^{2}$
and Baojiu Li$^{1}$
\\
$^{1}$Institute for Computational Cosmology, Department of Physics, Durham University, Durham DH1 3FE, United Kingdom\\
$^{2}$Department of Computer Science, Durham University, Durham DH1 3FE, United Kingdom\\
$^{3}$Institute for Data Science, Large-Scale Computing, Durham University, Durham DH1 3FE, United Kingdom\\
}

\date{Accepted XXX. Received YYY; in original form ZZZ}

\pubyear{2022}

\begin{document}
\label{firstpage}
\pagerange{\pageref{firstpage}--\pageref{lastpage}}
\maketitle

\begin{abstract}
The spherical collapse scenario has great importance in cosmology,
since it captures several crucial aspects of structure formation.
The presence of self-similar solutions in the Einstein-de Sitter (EdS) model greatly simplifies its analysis, making it a powerful tool to gain valuable insights into the real and more complicated
physical processes involved in galaxy formation. While there has been a large body of research to incorporate various additional physical processes into spherical collapse, the effect of modified gravity (MG) models, which are popular alternatives to the $\Lambda$CDM paradigm to explain the cosmic acceleration, is still not well understood in this scenario. In this paper, we study the spherical accretion of collisional gas in a particular MG model, which is a rare case that also admits self-similar solutions. The model displays interesting behaviours caused by the enhanced gravity and a screening mechanism. Despite the strong effects of MG, we find that its self-similar solution agrees well with that of the EdS model. These results are used to assess a new cosmological hydrodynamical code for spherical collapse simulations introduced here, which is based on the hyperbolic partial differential equation engine \exahype{} 2.
Its good agreement with the theoretical predictions confirms the reliability of this code in modelling astrophysical processes in spherical collapse. We will use this code to study the evolution of gas in more realistic MG models in future work.
\end{abstract}

\begin{keywords}
Numerical hydrodynamics -- Modified gravity -- Spherical collapse
\end{keywords}



\input{Introduction}

\input{Theory}

\input{exahype}

\input{result_and_conclusion}

\input{acknowledgements}

\section*{Data Availability}

\exahype{} 2 is available as part of the Peano 4 adaptive mesh refinement framework (\url{www.peano-framework.org}). For access to other data described by this paper, please contact HZ.


\bibliographystyle{mnras}
\bibliography{main} 


\appendix

\section{spherical symmetry of solution}
\label{app:spherical_symmetry}

\begin{figure*}
    \centering
    \includegraphics[width=\textwidth]{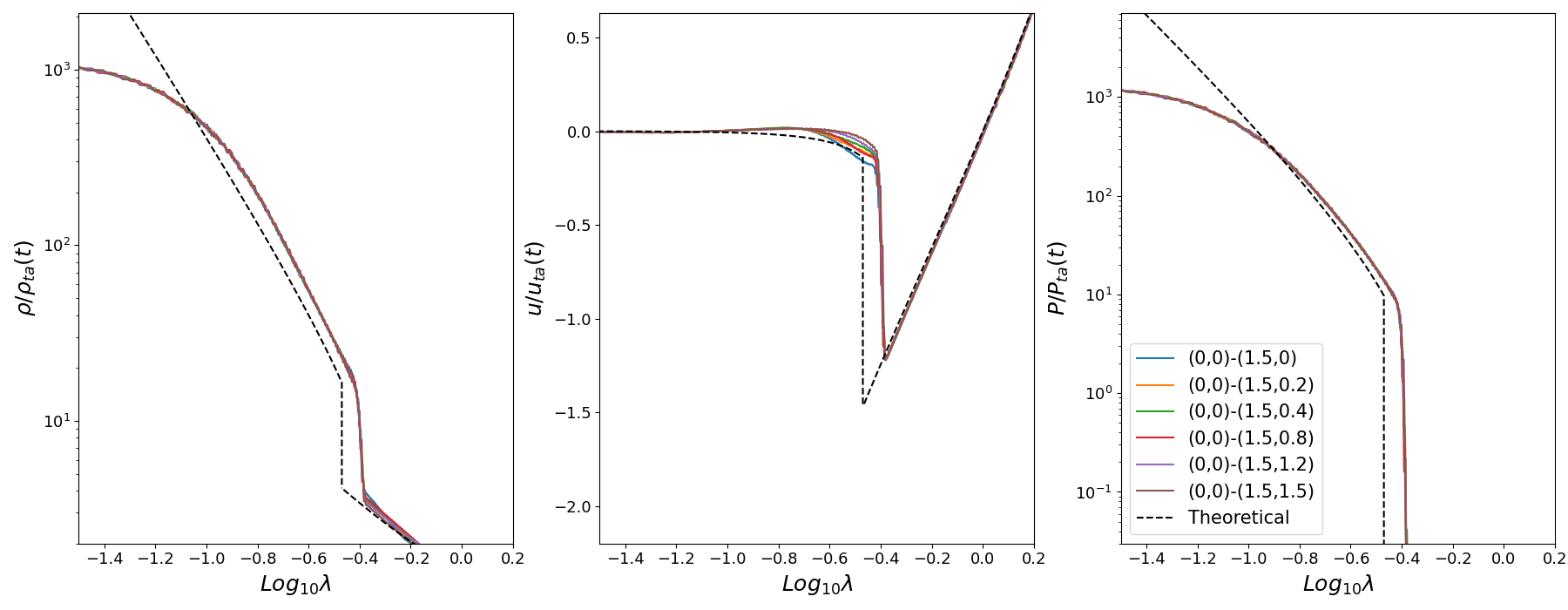}
    \caption{(Colour Online) The rescaled profiles of physical quantities in six different directions (as given in the legend) from
    the same simulation. It shows that the profiles of all considered quantities only have a very weak dependence on the direction along which we sample the solutions. See the text for more details.}
    \label{fig:sph_sym}
\end{figure*}

In this appendix, we show that our simulation is highly close to exact spherical symmetry during the stable evolution phase. In Figure \ref{fig:sph_sym}, we plot the rescaled profiles of the same three physical quantities discussed in Section \ref{sec4}, sampled along
six different directions (shown in different colours) from the same simulation. The sampling directions are all on the $x$-$y$ plane from a slice of the simulation box perpendicular to the $z$-axis. The coordinates shown in the legend are the starting 
and ending points of the sampling axis, 
while the black dashed line is the theoretical self-similar prediction plotted for comparison. 

These profiles all agree with each other nearly perfectly except in the small region immediately inside the shock in the velocity profile (middle) panel, where the curves in difference directions deviate from each other slightly and the profile in the diagonal 
direction (brown line, $(0,0)$--$(1.5,1.5)$) shows the most similar shape to the theoretical pattern (though the simulation result has a different amplitude due the reason explained in the main text). In particular, we note that the shock position is in good agreements along the different directions.

\bsp	
\label{lastpage}
\end{document}

%% file: Introduction.tex
\section{Introduction}

Spherical collapse is a widely studied phenomenon in cosmology.  
It describes the evolution of a spherically symmetric overdense region: 
how it slows down and decouples from the Hubble flow, turns around, and finally collapses into a singularity or some virialised matter distribution. Despite its simplicity, this scenario 
is of great importance, as it 
can describe several crucial aspects of structure formation of different matter components (e.g., collisionless dark matter and collisional baryonic gas), 
thus providing valuable insights 
into the real and more complicated cosmological process. 
Some cosmological hydrodynamical simulation codes also adopt this scenario as a test of reliability \citep[e.g., \ramses,][]{ramses_paper}.

The study of spherical collapse has a long history, with some of the early works including \citet[][]{Gunn1972,fillmore84,ryden87,Subramanian_2000}. 
Among them, \cite{1985ApJS...58...39B}
revealed an elegant self-similarity in the solution for a matter-dominated, Einstein-de Sitter (EdS), universe, 
for both collisionless and collisional matter. 
Using the turnaround radius, 
$r_{\rm ta}(t)$, in the EdS model, 
the various quantities in the system of evolution equations can be rescaled, such that all the dependencies on the spherical radius $r$ and time $t$ are reduced into the dependence on a single variable $\lambda{\equiv}r/r_{\rm ta}(t)$. This gives a unique set of solutions of physical quantities, expressed in terms of $\lambda$, which can be used to obtain the status of the evolution at arbitrary $(r,t)$.
Spherical collapse is therefore one of the few scenarios where a detailed semi-analytical solution is known in cosmology.

In the past decades, a lot of effort has been made to
incorporate more physical processes
into the spherical collapse model. Based on original radial collapse of matter, there are 
studies that look into the effects of angular momentum \citep{ryden88,sikivie97,delliou03}, 
dynamical friction \citep{antonucci94,Del_Popolo_2009} and shears \citep{popolp13,Pace14}. On the thermodynamics side, some research also studies cooling and heating process during the collapse \citep{abadi00,uchida04,McCarthy07} for a more realistic thermal history of structure formation.

Nowadays, studies of cosmological structure formation have entered a highly advanced stage, with ever more complicated physical processes added into increasingly sophisticated hydrodynamical simulations \citep[e,g,][]{Schaye:2014tpa,McCarthy:2016mry,Springel:2017tpz}, which can realistically reproduce the observed properties of galaxies and clusters;
see, e.g., \citet{borgani11}, for a review. Comparatively, therefore, the role of spherical collapse as a stand-alone simulation experiment has declined. 
However, this scenario can still be served as an useful benchmark test to access the accuracy and reliability of new simulation codes.
This explains partially why the original spherical collapse model with self-similarity still
attracts attention \citep[e.g.,][]{halle2019,alard20}. It is also worth noting that the self-similarity can still hold under some other circumstances, if the physics added (e.g., the cooling function) follows certain assumptions \citep{sikivie97,uchida04}. 

This paper concerns spherical collapse in modified gravity (MG) models, which are an alternative solution to avoid several problems of the current concordance $\Lambda$CDM cosmological model. The $\Lambda$CDM model suggests that the majority of 
energy density in the Universe is cold dark matter (CDM), 
a species of non-baryonic and non-realistic particles, and dark energy, an energy component with exotic properties (e.g., negative pressure), in the form of a positive cosmological constant $\Lambda$ \citep[][]{Amendola10}, which is needed to explain the accelerated Hubble expansion. 
However, the hypothetical $\Lambda$ suffers from long-standing theoretical problems 
\citep{weinberg89}. Modified gravity models hope to overcome those issues by extending the standard General Relativity (GR) rather than assuming extra unobserved components \citep[e.g.,][]{fR10,fT10}. 
In recent years, there has been growing interest in the MG models, because constraining them in various astrophysical and cosmological observations offers a powerful way to test our theory of gravity.

In this context, we want to mention that there are already 
various studies which look into the spherical collapse scenarios in different modified gravity models \citep[e.g.,][]{martino09,2010PhRvD..81f3005S,2012MNRAS.421.1431L,Lombriser:2013eza,Barreira:2014zza,Lopes_2018,contigiani19}. However, these studies generally focus on models that no longer uphold the property of self-similarity. This is not surprising, because even within GR there are strict conditions which must be satisfied to have self-similar solutions. For example, the EdS model loses its self-similarity property once a cosmological constant is added.

In this paper, we investigate the spherical collapse scenario for collisional gas in both Einstein-de Sitter universe and a slightly modified version of the Dvali-Gabadadze-Porrati \citep[DGP, ][]{dvali00}
braneworld model. The latter is a popular class of MG models that has attracted much attention in the last two decades, featuring an enhanced strength of the total gravitational force and the Vainshtein screening mechanism \citep[][]{Vainshtein:1972sx} which suppresses deviations from GR near massive objects to give the model a chance of passing the stringent Solar System and lab constraints. Despite its complexity, we find that the self-similarity property can still be achieved in 
this model under 
certain conditions which are not unnatural.
Our self-similar solutions in this model will provide insights into how these mechanisms of modified gravity may affect 
structure formation in similar, but more
realistic models where self-similarity no longer happens. 

We then implement the spherical collapse scenario in 
a new cosmological hydrodynamical code, and use the above derived self-similar solutions to assess its ability in handling 
simulations for different gravity models. Our code is based upon the publicly-available hyperbolic partial differential equation (PDE) engine \exahype{} 2, which implements a blockstructured adaptive mesh refinement (AMR) \citep{Dubey:16:SAMR} Finite Volume \citep{leveque_2002} code  on spacetrees \citep{Weinzierl:19:Peano}, and is parallelised through a combination of MPI, OpenMP BSP parallelism and a task formulation \citep{Li:22:ISC}. A comparison of the theoretical solutions and simulation results proves the reliability of our code, and we are going to use it to study more realistic and complicated modified gravity models in future work.

Our paper is organised as follows: in Section \ref{sec2} we briefly introduce the DGP model (\S~\ref{subsect:dgp_model}), and review the self-similar solution in an Einstein-de Sitter universe (\S~\ref{subsubsect:ss_solns_eds}) discovered by \citet{1985ApJS...58...39B}.
We then derive the self-similar solution for a slightly modified version of the DGP model, following a similar approach (\S~\ref{subsubsect:dgp_spherical}), and compare the behaviour of the solutions in this model, for several difference parameter choices, with that of the EdS model (\S~\ref{subsect:ss_solns}). This analysis reveals some interesting features of the solutions, which will be discussed in detail there. In Section \ref{sec3} we describe our numerical code and the simulation configuration we use for the spherical collapse scenario, paying particular attention to the implementation details and certain tricky issues in the settings including the initial and boundary conditions (\S~\ref{subsect:sim_settings}). The simulation results are presented and discussed in Section \ref{sec4}, and finally Section \ref{sec5} is devoted to discussions and conclusions.

%% file: Theory.tex
\section{Theories}
\label{sec2}

In this section, we introduce the physics we investigated and implemented in the code. 

Throughout this paper, we assume that the background cosmology is that of the Einstein-de Sitter
universe, 
i.e., a flat matter-dominated background(for simplicity we assume that this still holds even in the DGP models). It used to be the standard 
cosmological model before the $\Lambda {\rm CDM}$ model 
replaced it in the face of growing evidence that the cosmic expansion rate has been accelerating at late times, and it 
still serves as a good approximation for the real Universe 
between redshifts $\simeq300$ and $\simeq2$. The EdS universe assumes a zero cosmological constant and flat spatial curvature, and the equation of state of its non-relativistic matter content is $P(\rho)=0$. 
With these parameters, the evolution of the scale factor of the universe, $a$, can be derived analytically from the Friedmann equation as $a(t)={C}t^{2/3}$, where $t$ is the cosmic time, $C\equiv t_0^{-2/3}$ is a constant and $t_0$ is the cosmic time today (when $a=1$). This is an important assumption we will use to derive the self-similar solution later. 

\subsection{The DGP gravity model} 
\label{subsect:dgp_model}

The Dvali-Gabadadze-Porrati (DGP) braneworld model is a modified gravity model in a spacetime with an extra, fifth, dimension. The base assumption of this 
model 
is that the universe is a four-dimensional ``brane'' embedded in a
five-dimensional spacetime, which is called a ``bulk''.

This  
model provides an explanation as why gravity is much weaker than other fundamental forces: all matter components are assumed to be confined on the brane, while gravitons could propagate through, or leak into, the extra spatial dimension. 

The spacetime action of the DGP model is given by
\begin{align}
    S = \int_{\text{brane}} {\rm d}^4{x} \sqrt{-g} \frac{R}{16 \pi G} + \int_{\text{bulk}} {\rm d}^5{x} \sqrt{-g^{(5)}} \frac{R^{(5)}}{16 \pi G^{(5)}}, \label{eqn:grav_act_DGP}
\end{align}
where $R$ is the Ricci scalar, $g$ is the determinant of the metric tensor, $G$ is Newton's constant, and the superscript $^{(5)}$ means the corresponding quantities live in the five-dimensional bulk. Others without it are normal four-dimensional quantities.  

The modified Einstein equation for the DGP models can be derived
from the variation of the gravitational action Eq.~(\ref{eqn:grav_act_DGP}), which further leads to the following modified Friedmann equation that governs the cosmic expansion history $H(a)$:
\begin{equation}
    \frac{H(a)}{H_0} = \sqrt{\Omega_{\rm m0} a^{-3} + \Omega_{\rm DE}(a) + \Omega_{\rm rc}} \pm \sqrt{\Omega_{\rm rc}} , \label{eqn:Friedmann_eqn_nDGP}
\end{equation}
where $H_0=H(a=1)$ is the Hubble constant today (when the scale factor is $a=1$), $\Omega_{\rm m0}$ is the present-day density parameter of matter (we have neglected the presence of radiation and massive neutrinos here since they are not relevant for the interest of this work), $\Omega_{\rm DE}(a)$ represents the density parameter of a possible additional dark energy species at time $a$, and $\Omega_{\rm rc} \equiv c^2 / (4H_0^2 r_c^2)$. Here, $r_c$ is the so-called crossover scale, 
which is a new free model parameter that indicates the scale above which the gravity begins to deviate from the standard Einsteinian:
\begin{equation}
    r_c \equiv \frac{1}{2} \frac{G^{(5)}}{G}.
\end{equation}
It is easy to see that, Eq.~(\ref{eqn:Friedmann_eqn_nDGP}) goes back to the usual form of the Friedmann equation when $H_0 r_c \to \infty$.

The $\pm $ in Eq.~\eqref{eqn:Friedmann_eqn_nDGP} shows that this
model has two branches of solutions. There is a ``self-accelerating'' branch (sDGP, the ``$+$'' branch) that can realise an accelerated Hubble expansion 
at late times without the need of a cosmological constant or dark energy, i.e., $\Omega_{\rm DE}(a)=0$. However, this branch has several unsolved theoretical issues \citep{2007CQGra..24R.231K}. Additionally, its predicted cosmological history is significantly different from that of $\Lambda{\rm CDM}$ and the observation also disfavours this model \citep[e.g.,][]{2007PhRvD..75f4003S}.

The other branch, the so-called normal branch of DGP (nDGP) gravity, where Eq.~\eqref{eqn:Friedmann_eqn_nDGP} takes the ``$-$'' sign, can not provide an accelerated Hubble expansion by itself, and thus some additional dark energy component is needed ($\Omega_{\rm DE}\neq0$) to explain the observation. 
This model has attracted much attention
in recent years as it serves as a useful testbed of the Vainshtein screening mechanism \citep[e.g.,][]{Brax13}, despite its unappealing property of being still in need for additional dark energy. We will describe Vainshtein screening in more detail below. 

The (modified) Poisson equation of DGP gravity and corresponding equations of
the scalar field have been derived by \citet{koyama07prd}:
\begin{equation}\label{eq:poisson_nDGP}
\nabla^2 \Phi = 4\pi G a^2 \delta \rho_{\rm m} + \frac{1}{2}\nabla^2\varphi\,,
\end{equation}
and
\begin{equation}\label{eq:phi_dgp}
\nabla^2 \varphi + \frac{r_c^2}{3\beta\,a^2c^2} \left[ (\nabla^2\varphi)^2
- (\nabla_i\nabla_j\varphi)^2 \right] = \frac{8\pi\,G\,a^2}{3\beta} \delta\rho_{\rm m},
\end{equation}
where $\Phi$ and $\varphi$ are 
the gravitational
potential and the scalar field of the model, respectively. They are
also known as the brane-bending mode, which represents the position of the brane in the fifth dimension. $\nabla$ is the spatial gradient (wrt to comoving coordinates), $c$ is the speed of light and $\delta\rho_{\rm m} = \rho_{\rm m} - \bar{\rho}_{\rm m}$ is the matter density perturbation (throughout this paper an overbar denotes the background value of a quantity).  $\beta$ is a time-dependent function:
\begin{equation}\label{eq:beta_dgp}
\beta(a) \equiv 1 \pm 2 H\, r_c \left ( 1 + \frac{\dot H}{3 H^2} \right ),
\end{equation}
for the two branches, which for the normal branch can be simplified as
\begin{equation}
    \beta(a) = 1 + \frac{\Omega_{\rm m0}a^{-3} + 2\Omega_{\rm rc}}{2\sqrt{\Omega_{\rm rc}(\Omega_{\rm m0}a^{-3} + \Omega_{\rm rc})}}.
\end{equation}

While we are interested in the DGP model, our main focus in this paper will be the effect of a \textit{fifth force} that is mediated by the scalar field $\varphi$, denoted by the second term on the right-hand side of Eq.~\eqref{eq:poisson_nDGP}. To gain flexibility and to ensure self-similarity of the resulting model behaviour, we take the liberty to keep the main features of Eq.~\eqref{eq:phi_dgp} but allow deviations from the exact behaviour of the sDGP or nDGP models. More explicitly, we will promote $r_c$ to a time-dependent function, and also allow $\beta$ to differ from Eq.~\eqref{eq:beta_dgp}. We remark that such variations from the original DGP model are not uncommon in other modified gravity models involving the Vainshtein mechanism, notably the cubic Galileon \citep[][]{Nicolis:2008in,Deffayet:2009wt} and the Proca \citep[][]{Heisenberg:2014rta} theories.

\subsection{Self-similar behaviour in collapse of collisional gas}
\label{subsect:ss_derivation}

In this subsection, we describe the self-similar collapse of collisional and non-radiative gas in some models. We first review the classic result from \citet{1985ApJS...58...39B}, which applies to standard gravity in EdS universe. Then we proceed to show that self-similarity can also be achieved in the DGP model with Vainshtein screening. These can be used as a test case to verify our numerical implementation with \exahype{} 2 for both the standard and modified gravity scenarios, though our implementation of modified gravity is not restricted to the DGP model where self-similarity holds.

\subsubsection{Einstein-de Sitter universe}
\label{subsubsect:ss_solns_eds}

Consider a uniform spherical overdensity region in the matter dominated universe background. Its initial condition could be written as
\begin{equation}
    \rho=\frac{1}{ 6\pi G t^2_i}\begin{cases}
    1+\delta_i,&\quad r > R_i\\
    1,&\quad r < R_i,
    \end{cases}
    \label{eq::initial_density}
\end{equation}
$t_i<t_0$ is the initial cosmic time for this scenario to begin, $\delta_i=\delta \rho/\bar{\rho}\ll 1$ the density contrast at $t_i$, where $\bar{\rho}=\bar{\rho}(t)$ and $\delta\rho$ are respectively the mean matter density at time $t$ and the density perturbation, and $R_i$ is the initial radius of the spherical overdensity region. At the beginning, the Hubble flow is approximately unperturbed as $\delta_i\ll1$. Thus, we have $v_i=H_i r_i$ and $H_i=2/(3t_i)$. As the universe expands, the matter inside $R_i$ starts to decelerate and  decouple from the Hubble flow because of the slightly higher density. At some point it stops expanding completely (so-called ``turnaround'') and turns into a collapse. The turnaround for the mass shell at $R_i$ initially happens at a cosmic time and max radius \citep{1985ApJS...58...39B}
\begin{equation}
    t_{\rm ita} = \frac{3\pi}{4}\delta_i^{-3/2}t_i, \quad
    r_{\rm ita} = R_i \delta_i^{-1},
    \label{eq::initial_turnaround_time_radius}
\end{equation}
where the subscript $_{\rm ita}$ stands for ``initial turnaround''. Matter inside the initial overdensity region starts to collapse first and all matter there infalls at the same time. No shell crossing happens. The matter initially in more distant shells (i.e., at initial radii $r_i>R_i$) will start to collapse in progressively later times. The radius at which they turn around can be calculated using the Lagrangian picture. For the mass element initially located at $r_i$, its evolution obeys the Newton's gravity law:
\begin{equation}
    \frac{{\rm d}^2 r}{{\rm d} t^2}=-\frac{G m}{r^2}.
    \label{eq::newton_gravity}
\end{equation}
Here $m$ accounts all mass interior to the shell we are considering. As no shell crossing happens during the evolution, it can be written as
\begin{equation}\label{eq::mass_Delta}
    m = m\left(r_i\right) = \frac{4}{3}\pi\rho_ir_i^3\left(1+\delta_i\frac{R^3_i}{r_i^3}\right) \equiv \frac{4}{3}\pi\rho_ir^3_i(1+\Delta),
\end{equation}
where $\rho_i=1/(6\pi G t_i^2)$ is the background density at $t_i$ for an Einstein-de Sitter universe. We then recast equation (\ref{eq::newton_gravity}) using the following dimensionless time and radius variables, $\tau\equiv t/ t_i$ and $y\equiv r/ r_i$, as
\begin{equation}\label{eq:rescaled_newton_gravity}
    \frac{{\rm d}^2 y}{{\rm d} \tau^2} = -\frac{2}{9}\left(1+\Delta\right)\frac{1}{y^2}.
\end{equation}

Integrating this equation twice and using the assumption $\Delta\ll1$, the solution can be expressed implicitly as \citep[][]{1985ApJS...58...39B}
\begin{equation}\label{eq:soln_implicit_1}
    \tau = \frac{3}{4}\left(\theta-\sin\theta\right)\Delta^{-3/2} \equiv d\Delta^{-3/2},
\end{equation}
with 
\begin{equation}\label{eq:soln_implicit_2}
    y\Delta = \sin^2\frac{\theta}{2} \equiv \eta,
\end{equation}
where we have defined the variables $d$ and $\eta$ for later use. As turnaround happens when $y$ reaches its maximum, this yields to $\theta_{\rm ta}=\pi$ (where a subscript $_{\rm ta}$ means ``turnaround''). From Eq.~\eqref{eq:soln_implicit_1} this corresponds to a time
$\tau = (3\pi/4)\Delta^{-3/2}$ and $y_{\rm ta}=r_{\rm ta}/r_i=\Delta^{-1}$. Combining these two expressions with the relationship between $\delta_i$, $\Delta$ and $R_i$ given in Eq.~\eqref{eq::mass_Delta}, it is straightforward to derive the following expression of the turnaround radius:
\begin{equation}\label{eq::turn_around_radius}
    r_{\rm ta}(t) = \left(\frac{3\pi}{4}t_i\right)^{-8/9}\delta_i^{1/3}R_it^{8/9},
    \quad {\rm for} ~ t \geq t_{\rm ita}.
\end{equation}

Now we switch to the fluid picture. The motion of a collisional gas in this system is governed by the gravity-driven Euler equations:
\begin{eqnarray}
    \label{eq::gas_eq1}\frac{{\rm d}\rho}{{\rm d}t} &\equiv& \left[\frac{\partial}{\partial{t}}+v\frac{\partial}{\partial{r}}\right]\rho = -\rho\frac{1}{r^2}\frac{\partial}{\partial{r}}\left(r^2v\right),\\
    \label{eq::gas_eq2}\frac{{\rm d}v}{{\rm d}t} &=& -\frac{1}{\rho}\frac{\partial{p}}{\partial{r}} - \frac{Gm}{r^2},\\
    \label{eq::gas_eq3}\frac{{\rm d}}{{\rm d}t}\left(p\rho^{-\gamma}\right) &=& 0,\\
    \label{eq::gas_eq4}\frac{\partial m}{\partial r} &=& 4\pi{r}^2\rho,
\end{eqnarray}
where $\rho=\rho(r,t)$, $v=v(r,t)$ and $p=p(r,t)$ are the density, velocity and pressure of the fluid at radius $r$ and time $t$. $m\equiv m(<r)$ represents the total mass within a given radius $r$, and $\gamma$ the adiabatic index. We now use Eq.~(\ref{eq::turn_around_radius}) and define the new radial coordinate:
\begin{equation}
    \label{eq::lambda}
    \lambda\equiv \frac{r}{r_{\rm ta}},
\end{equation}
as well as the dimensionless quantities $V, D, P$ and $M$:
\begin{eqnarray}
    \label{eq:dimensionless_v}v(r,t) &=& \frac{r_{\rm ta}}{t}V(\lambda),\\
    \label{eq:dimensionless_r}\rho(r,t) &=& \rho_{\rm H}D(\lambda),\\
    \label{eq:dimensionless_p}p(r,t) &=& \rho_{\rm H}\left(\frac{r_{\rm ta}}{t}\right)^2P(\lambda),\\
    \label{eq:dimensionless_m}m(r,t) &=& \frac{4\pi}{3}\rho_{\rm H}r^3_{\rm ta}M(\lambda),
\end{eqnarray}
where $\rho_{\rm H}=\rho_{\rm H}(t)$ is the critical density at time $t$, which is equal to the mean matter density $\bar{\rho}_{\rm m}(t)$ in the EdS model. These allow us to cast Eq.~(\ref{eq::gas_eq1} - \ref{eq::gas_eq4}) as the following new dimensionless fluid equations \citep{1985ApJS...58...39B}:
\begin{eqnarray}
    \label{eq:eq1_dimensionless}\left(V-\frac{8}{9}\lambda\right)D' + DV' + 2\frac{D}{\lambda}V - 2D &=& 0,\\
    \label{eq:eq2_dimensionless}\left(V-\frac{8}{9}\lambda\right)V' - \frac{1}{9}V &=& -\frac{P'}{D} - \frac{2}{9}\frac{M}{\lambda^2},\\
    \label{eq:eq3_dimensionless}\left(V-\frac{8}{9}\lambda\right)\left(\frac{P'}{P}-\gamma\frac{D'}{D}\right) &=& \frac{20}{9}-2\gamma,\\
    \label{eq:eq4_dimensionless}M' &=& 3\lambda^2D,
\end{eqnarray}
where a prime means the derivative wrt $\lambda$. Those equations only have one variable $\lambda$ and thus could be solved directly given proper boundary conditions (see Section \ref{subsect:ss_solns} below). No further time or length scales are involved, which means that the solutions to the system would remain identical throughout the evolution if expressed in terms of the $\lambda$ coordinate. This is where self-similarity comes from. Obviously, if any new terms added in Eqs. (\ref{eq::gas_eq1} - \ref{eq::gas_eq4}) depend on other scales besides $\lambda$, the solution to that new system will deviate from this self-similar solution.

\subsubsection{DGP gravity model}
\label{subsubsect:dgp_spherical}

In the spherically symmetric system, the scalar field equation (\ref{eq:phi_dgp}) gets simplified significantly \citep[see, e.g.,][]{baojiu13}:
\begin{equation}\label{eq:scalar_in_spherical}
    \frac{2r_c^2}{3\beta{c}^2}\frac{1}{r^2}\frac{\partial}{\partial r}\left[r \left(\frac{\partial\varphi}{\partial r}\right)^2 \right] + \frac{1}{r^2} \left[r^2 \frac{\partial\varphi}{\partial r} \right]=\frac{8\pi G}{3\beta}\delta \rho_m.
\end{equation}
This equation does not contain scale factor $a$ as we use physics radius here. We then define
\begin{equation}\label{eq:m_hat_def}
    \hat{m}(r)\equiv 4\pi\int^r_0 \delta \rho_m(r')r'^2{\rm d}r',
\end{equation}
where we have used $\hat{m}$ to distinguish from $m(r)$ introduced in Eq.~\eqref{eq::newton_gravity}, since $\hat{m}$ does not account for the background matter density. Eq.~(\ref{eq:scalar_in_spherical}) then can be integrated once to give:
\begin{equation}\label{eq:vainshtein_eqn_spherical}
    \frac{2r_c^2}{3\beta{c}^2}\frac{1}{r}\left(\frac{\partial\varphi}{\partial{r}}\right)^2 + \frac{\partial\varphi}{\partial{r}} =\frac{2}{3\beta} \frac{G\hat{m}(r)}{r^2} \equiv \frac{2}{3\beta}g_{\rm N}(r),
\end{equation}
solving which gives the radial gradient of scalar field directly as
\begin{equation}\label{eq:Vainshtein_soln1}
    \frac{\partial\varphi}{\partial{r}} = \left[2\frac{2r_c^2}{3\beta{c}^2}\frac{1}{r}\right]^{-1}\left[-1+\sqrt{1+4\frac{2r_c^2}{3\beta{c}^2}\frac{1}{r}\frac{2}{3\beta}g_{\rm N}}\right], 
\end{equation}
where we have dropped the other branch of solution that is unphysical. The equation could be simplified further by defining the ``Vainshtein radius'' $r_V$ as follows:
\begin{equation}\label{eq:Vainshtein_radius}
    r_V^3 = \frac{16r_c^2}{9\beta^2c^2}G \hat{m}\left(r_{\rm ta}\right).
\end{equation}
Note that here we have used $\hat{m}$ within $r_{\rm ta}(t)$ to define the Vainshtein radius, which differs from the usual definition that only accounts for the mass within the tophat radius --- this is for convenience, because in this way we end up with a generic expression that does not depend on the particular size of any tophat. Now the gradient of scalar field reads as:
\begin{equation}\label{eq:Vainshtein_soln}
    \frac{\partial\varphi}{\partial{r}} = \frac{4}{3\beta}\frac{r^3}{r_V^3}\frac{\hat{m}\left(r_{\rm ta}\right)}{\hat{m}(r,t)}\left[\sqrt{1+\frac{r^3_V}{r^3}\frac{\hat{m}(r,t)}{\hat{m}\left(r_{\rm ta}\right)}}-1\right]g_{\rm N}(r).
\end{equation}
Note that $\partial \varphi /\partial r$ determines the strength of the fifth force, and one can easily see the following limiting behaviour:
\begin{eqnarray}
    \frac{\partial\varphi}{\partial{r}} \simeq \frac{2}{3\beta}g_{\rm N}(r), && r\gg r_V,\nonumber\\
    \frac{\partial\varphi}{\partial{r}} \ll \frac{2}{3\beta}g_{\rm N}(r), && r\ll r_V.
\end{eqnarray}
If the scale of studied problem is significantly smaller than the Vainshtein radius $r_V$, the gradient of the scalar field is also much smaller than that of the Newtonian potential, such that the fifth force is negligible compared with the standard Newtonian force. This is the idea behind the Vainshtein screening.

Our next step is to try to recast the expression of the fifth force in the self-similar form (which, needless to say, is not always possible) similar to what we get above for the Einstein-de Sitter universe. This means that we hope that the ratio between the fifth force and standard Newtonian gravity, i.e., the coefficient in front of Eq.~\eqref{eq:Vainshtein_soln}, depends on time $t$ and radius $r$ only through the combination $r_{\rm ta}(t)$. We again define $\lambda\equiv r/r_{\rm ta}$; note that this $r_{\rm ta}$
is the same as in Eq.~(\ref{eq::turn_around_radius})--- this is mainly for convenience, but it does mean the $r_{\rm ta}$ in this expression is no longer the true turnaround radius in the DGP model. The mass can then be rewritten, using the definition of $M(\lambda)$ give in Eq.~\eqref{eq:dimensionless_m}, as
\begin{equation}
    \hat{m}(r,t) = \frac{4\pi}{3}\rho_{\rm H}r^3_{\rm ta}\left[M(\lambda)-\lambda^3\right] \equiv \frac{4\pi}{3}\rho_{\rm H}r^3_{\rm ta}\hat{M}(\lambda).
\end{equation}
As mentioned above, we have removed the contribution from the background mass as the fifth force only depends on density perturbations. The Vainshtein radius now reads as
\begin{equation}
    r^3_V = \frac{16r_c^2}{9\beta^2c^2}\frac{2}{9}\frac{r^3_{\rm ta}}{t^2}\hat{M}(1),
\end{equation}
so that
\begin{equation}
    \frac{r_V^3}{r^3} = \frac{16r_c^2}{9\beta^2c^2}\frac{2}{9t^2}\frac{\hat{M}(1)}{\lambda^3},
\end{equation}
and
\begin{equation}
    \frac{\hat{m}(r,t)}{\hat{m}\left(r_{\rm ta}\right)} = \frac{\hat{M}(\lambda)}{\hat{M}(1)}. 
\end{equation}
These mean that
\begin{equation}\label{eq:Vainshtein_soln_key_term}
    \frac{r_V^3}{r^3}\frac{\hat{m}(r,t)}{\hat{m}\left(r_{\rm ta}\right)} = \frac{16r_c^2}{9\beta^2c^2}\frac{2}{9t^2}\frac{\hat{M}(\lambda)}{\lambda^3}.
\end{equation}
To achieve the self-similarity, we need to ensure that Eq.~(\ref{eq:Vainshtein_soln_key_term}) only depends on $\lambda$. The $t$ dependence of $r^2_c/\beta^2 t^2$ need to be cancelled out. However, $\beta$ also appears in Eq.~(\ref{eq:Vainshtein_soln}) in the overall factor $4/(3\beta)$,
and thus should be constant over time to avoid reintroducing an explicit $t$ dependency.
This then leads to $r_c\propto{t}\propto{a}^{3/2}$, with the second proportionality true in an Einstein-de Sitter universe.

Denoting $r_c(t)=r_{c0}(t/t_0)$, where $t_0$ is the cosmic time today and $r_{c0}$ is the value of $r_c$ at $t_0$, and defining the dimensionless constant
\begin{equation}\label{eq:zeta_def}
    \zeta \equiv \frac{r_{c0}}{ct_0} = \frac{r_{c0}\times\left(t/t_0\right)}{ct} =\frac{r_c(t)}{ct}= \frac{3H(t)r_c(t)}{2c} = \frac{3H_0r_{c0}}{2c},
\end{equation} 
we get
\begin{equation}\label{eq:Vainshtein_soln_key_term_with_zeta}
    \frac{r_V^3}{r^3}\frac{\hat{m}(r,t)}{\hat{m}\left(r_{\rm ta}\right)} = \frac{32\zeta^2}{81\beta^2}\frac{\hat{M}(\lambda)}{\lambda^3}.
\end{equation}
Therefore, the solution can be written as
\begin{equation}\label{eq:dphidr_1}
    \frac{\partial\varphi}{\partial{r}} = \frac{27\beta}{8\zeta^2}\frac{\lambda^3}{\hat{M}(\lambda)}\left[\sqrt{1+\frac{32\zeta^2}{81\beta^2}\frac{\hat{M}(\lambda)}{\lambda^3}}-1\right]g_{\rm N}(r).
\end{equation}
This expression shows that the fifth-force-to-Newtonian-gravity ratio can be written in a form that only depends on $\lambda$, which satisfies the requirement of self-similarity. It is straightforward to show that the coefficient of $g_{\rm N}$ in the above equation is always smaller than $2/(3\beta)$, which means that the Vainshtein screening always works (though not necessarily always strong).

Let us briefly comment that, according to its definition in Eq.~\eqref{eq:zeta_def}, $\zeta$ is the ratio between the crossover radius $r_c(t)$ and $ct$. The latter can be considered as some characterisation of the size of the Einstein-de Sitter universe (actually it is $3ct$). Therefore, the fact that this ratio is a constant in time implies that the Vainsthein screening mechanism is always effective on scales that correspond to a fixed fraction of the size of the universe, and therefore it should not be surprising that the self-similar properties of the EdS model have been preserved for this particular choice of $r_c(t)$. Since $r_c$ characterises the length scale beyond which gravity is modified in the DGP model, we expect that for any physically interesting scenario we need to have $\zeta\sim\mathcal{O}(1)$. The choice of $\zeta=2/3$, for example, corresponds to $H_0r_{c0}/c=1$, which leads to a similar Vainshtein screening efficiency to that for a typical parameter choice in studies of the nDGP model for the same value of $\beta$.

The actual strength of the fifth force is $\frac{1}{2} \frac{\partial\varphi} {\partial{r}}$, which means that the final expression for the fifth-force-to-Newtonian-gravity ratio is given by
\begin{equation}\label{eq:xi_def}
    \xi(\lambda) \equiv \frac{27\beta}{16\zeta^2}\frac{\lambda^3}{\hat{M}(\lambda)}\left[\sqrt{1+\frac{32\zeta^2}{81\beta^2}\frac{\hat{M}(\lambda)}{\lambda^3}}-1\right].
\end{equation}
Turning to the derivation of the self-similar equations in the DGP model, i.e., the counterparts of Eqs.~(\ref{eq:eq1_dimensionless} - \ref{eq:eq4_dimensionless}), it is evident that only Eq.~\eqref{eq:eq2_dimensionless} needs to be modified. It is the only place where the law of gravity enters the calculation. However, instead of simply multiplying the $-\frac{2}{9}\frac{M}{\lambda^2}$ by $1+\xi(\lambda)$, the correct final version of Eq.~\eqref{eq:eq2_dimensionless} is slightly more complicated. This is because $\xi(\lambda)$ is the ratio between the fifth force and $g_{\rm N}$, which itself does not receive any contribution from the background matter density, c.f., ~Eq.~\eqref{eq:vainshtein_eqn_spherical}. On the other hand, the term $-\frac{2}{9}\frac{M}{\lambda^2}$ contains contributions from the background matter. Taking this into account leads to the following DGP version of Eq.~\eqref{eq:eq2_dimensionless}:
\begin{equation}
    \label{eq:eq2_dimensionless_dgp1}\left(V-\frac{8}{9}\lambda\right)V' - \frac{1}{9}V = -\frac{P'}{D} - \frac{2}{9}\frac{M}{\lambda^2} - \frac{2}{9}\frac{\hat{M}}{\lambda^2}\xi(\lambda),
\end{equation}
or equivalently
\begin{equation}
    \label{eq:eq2_dimensionless_dgp2}\left(V-\frac{8}{9}\lambda\right)V' - \frac{1}{9}V = -\frac{P'}{D} - \frac{2}{9}\frac{M}{\lambda^2}\left[1+\xi(\lambda)\right] + \frac{2}{9}\lambda\xi(\lambda).
\end{equation}

A similar modification also appears in the DGP counterpart of Eq.~(\ref{eq::newton_gravity}), which now reads
\begin{equation}\label{eq:rescaled_DGP_gravity}
    \ddot{y} = -\frac{2}{9}\left(1+\Delta\right)\frac{1}{y^2}
    \left[1+\xi\left(y,\tau\right)\right] + \frac{2}{9}\frac{y}{\tau^2}\xi(y,\tau),
\end{equation}
where $\xi$ has been defined in Eq.~\eqref{eq:xi_def}, but is now expressed in terms of the dimensionless radius and time, $y$ and $\tau$. More explicitly:
\begin{equation}\label{eq:DGP_gravity_modifier}
    \xi = \frac{27\beta}{16\zeta^2}\frac{y^3}{(1+\Delta)\tau^2-y^3}\left[\sqrt{1+\frac{32\zeta^2}{81\beta^2}\left(\frac{1+\Delta}{y^3}\tau^2-1\right)}-1\right].    
\end{equation}
This equation is needed for the exact solution of our equations in the next section.

Before concluding this subsection, let us note that one limit of the DGP model arises from $\zeta\rightarrow0$, in which Eq.~\eqref{eq:dphidr_1} approaches 
\begin{equation}
    \frac{\partial\varphi}{\partial{r}} \rightarrow \frac{2}{3\beta}g_{\rm N}(r),
\end{equation}
and so the fifth-force-to-Newtonian-gravity ratio approximately becomes $1/(3\beta)$, which is the linear-regime (i.e., no screening) solution. This corresponds to a time- and scale-independent enhancement of Newton's constant by a factor of $1/(3\beta)$ since we are assuming $\beta$ to be a constant here.  

\subsection{Self-similar solutions}
\label{subsect:ss_solns}

Our next step is to find the exact solution to our self-similar equations Eqs.~(\ref{eq::gas_eq1} - \ref{eq::gas_eq4}): the profile of $D(\lambda),~V(\lambda),~P(\lambda)$ and $M(\lambda)$. 

At the beginning stage, the spherical collapse can be described by a pressureless infall. Outside the radius of the tophat, the inner spherical shells infall at a greater speed than the outer shells, meaning that there is no shell-crossing or squeezing. However, when the infall speed of a given shell increases to a point where it exceeds the sound speed $c_s$ of the fluid, the shell impacts upon the fluid element inside it before there is enough time for the latter to adjust. A discontinuity of fluid properties, such as velocity, pressure and density, then starts to arise there, which is known as a shock. The shock location is our primary quantity of interest when we validate the outcome of our simulation. We assume the shock happens at radius $r_s$ or $\lambda_s\equiv r_s/r_{\rm ta}$ (the subscript $_s$ means shock), where we can apply the Rankine–Hugoniot jumping conditions, written in dimensionless forms:
\begin{eqnarray}
    D_2V_2 = D_1V_1 + \left(D_2-D_1\right)V_s,&
    \label{eq:jump_cond1_dimensionless}\\
    D_2V_2^2 + P_2 = D_1V_1^2 + P_1 + \left(D_2V_2-D_1V_1\right)V_s,&
    \label{eq:jump_cond2_dimensionless}\\
    \begin{aligned}
    D_2V_2&\left(\frac{\gamma}{\gamma-1}\frac{P_2}{D_2}+\frac{1}{2}V_2^2\right) - D_1V_1\left(\frac{\gamma}{\gamma-1}\frac{P_1}{D_1}+\frac{1}{2}V_1^2\right)
     \\&= V_s\left[D_2\left(\frac{1}{\gamma-1}\frac{P_2}{D_2}+\frac{1}{2}V_2^2\right) - D_1\left(\frac{1}{\gamma-1}\frac{P_1}{D_1}+\frac{1}{2}V_1^2\right)\right].
    \label{eq:jump_cond3_dimensionless}
    \end{aligned}
\end{eqnarray}
Here, a subscript 1 or 2 is used to denote the preshock and postshock values of a quantity, respectively, and $V_s$ is the dimensionless speed of the shock position itself. Physically, the three jumping conditions represent the continuity of mass, momentum and energy across the shock.

\begin{figure*}
    \centering
    \includegraphics[width=\textwidth]{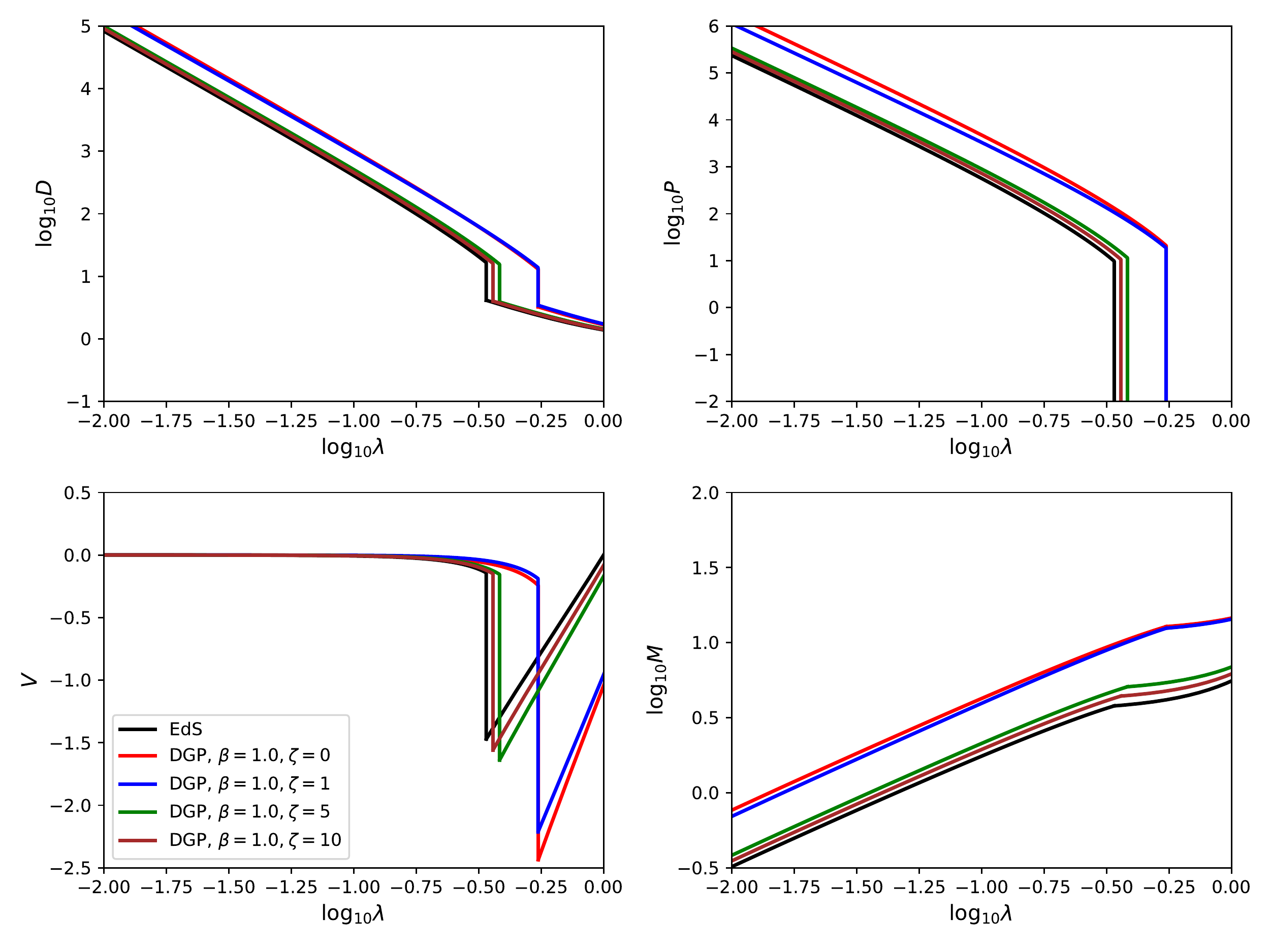}
    \caption{(Colour Online) Self-similar solution for gravity in Einstein-de Sitter universe and DGP models with different $\zeta$ choices. Rescaled $D(\lambda)$, $P(\lambda)$, $V(\lambda)$, $M(\lambda)$ are plotted.  The value of $\zeta$ indicates the strength of the Vainshtein screening. It gets more efficient when $\zeta$ gets bigger. The case $\zeta=0$ means there is no screening, i.e., The modification of DGP are equivalent to a constant enhancement of gravity all the time. All curves here are obtained by using $\Delta=0.001$.}
    \label{fig:self-similar_unrescaled}
\end{figure*}

One can analytically calculate the preshock solutions in 
terms of $\lambda_s$ using Eq.~(\ref{eq:rescaled_newton_gravity}) and its solutions, Eqs.~(\ref{eq:soln_implicit_1}, \ref{eq:soln_implicit_2}) for $\Delta\ll 1$:
\begin{eqnarray}
    \label{eq:pre_shock_soln_D}D_1 &=& \frac{d_s^2\eta_s^{-3}}{1+3\chi_s},\\
    \label{eq:pre_shock_soln_P}P_1 &=& 0,\\
    \label{eq:pre_shock_soln_V}V_1 &=& \lambda_s\frac{\sin\theta_s\left(\theta_s-\sin\theta_s\right)}{\left(1-\cos\theta_s\right)^2},\\
    \label{eq:pre_shock_soln_M}M_1 &=& \lambda^3_sd^2_s\eta_s^{-3}.
\end{eqnarray}
where $\theta_s=\theta\left(\tau_s\right)$,  $\eta_s \equiv \sin^2 \frac{\theta_s}{2}\equiv y_s\Delta$, $d_s \equiv \frac{3}{4} (\theta_s-\sin\theta_s)$ are the values of $\eta$ and $d$ at $\theta_s$, and, $\chi_s \equiv 1-\frac{3}{2}\frac{V_s}{\lambda_s}$. 
Combining Eqs.~(\ref{eq:jump_cond1_dimensionless} - \ref{eq:pre_shock_soln_M}), we get the boundary condition for the other side (post side) of the shock:
\begin{eqnarray}
    \label{eq:pst_shock_soln_D}D_2 &=& \frac{\gamma+1}{\gamma-1}D_1,\\
    \label{eq:pst_shock_soln_V}V_2 &=& \frac{8}{9}\lambda_s + \frac{\gamma-1}{\gamma+1}\left(V_1-\frac{8}{9}\lambda_s\right),\\
    \label{eq:pst_shock_soln_P}P_2 &=& \frac{2}{\gamma+1}D_1\left(V_1-\frac{8}{9}\lambda_s\right)^2,\\
    \label{eq:pst_shock_soln_M}M_2 &=& M_1.
\end{eqnarray}
The entire postshock solution can then be obtained by numerically integrating Eqs.~(\ref{eq:eq1_dimensionless} - \ref{eq:eq4_dimensionless}) inwards from $\lambda=\lambda_s$, using these boundary conditions. 
However, since $\lambda_s$ is not known \textit{a priori}, this is a trial and error process where the value of $\lambda_s$ is updated iteratively until when the corresponding solutions meet the following physical boundary conditions in the centre of the system:
\begin{equation}
    V(\lambda=0) = M(\lambda=0) = 0.
\end{equation}
This is how \cite{1985ApJS...58...39B} got his self-similar solution and we plot our reproduced result here in Figure \ref{fig:self-similar_unrescaled}.

While the use of the $\theta$ variable to write the solution to Eq.~\eqref{eq:rescaled_newton_gravity} in the implicit forms of Eqs.~(\ref{eq:soln_implicit_1}, \ref{eq:soln_implicit_2}) is convenient, this is impossible for the DGP model where the corresponding spherical collapse equation takes a more complicated form. However, the introduction of $\theta$ in the EdS model is largely a matter of choice for convenience, and the same physics can be produced using $\tau$ as well. Because this is what we shall use for the DGP model, we decide to also use $\tau$ instead of $\theta$ to obtain the numerical self-similar solutions for the EdS model. This means that we need to express the preshock solutions to $D, P, V$ and $M$ at $\tau_s$. For the velocity, using its definition
\begin{equation}
    v = \frac{{\rm d}r}{{\rm d}t} = \frac{r_i}{t_i}\dot{y} = \frac{r}{t}\frac{\tau}{y(\tau)}\dot{y} = \lambda\frac{r_{\rm ta}}{t}\frac{\tau}{y(\tau)}\dot{y}(\tau),\nonumber 
\end{equation}
we obtain
\begin{equation}\label{eq:dgp_ss_bc_v1}
     V_1(\lambda_s) = \lambda_s\frac{\tau_s}{y\left(\tau_s\right)}\dot{y}_s,
\end{equation}
where an overdot denotes the derivative wrt $\tau$, and $\dot{y}_s = \dot{y}\left(\tau_s\right)$. For $M$, using
\begin{equation}
    M=\frac{m}{\frac{4}{3}\rho_{\rm H}r_{\rm ta}^3}
    =\frac{\rho_i}{\rho_{\rm H}}\left(\frac{r_i}{r_{\rm ta}}\right)^3(1+\Delta),\nonumber
\end{equation}
we have 
\begin{equation}\label{eq:dgp_ss_bc_m1}
    M_1(\lambda_s) = \left(\frac{3\pi}{4}\right)^{8/3}\frac{1+\Delta}{\Delta}\tau_s^{-2/3}.
\end{equation}
For $D$, using
\begin{equation}
    3\lambda^2D(\lambda) = \frac{{\rm d}M/{\rm d}\tau}{{\rm d}\lambda/{\rm d}\tau} = \frac{-\frac{2}{3}\left(\frac{3\pi}{4}\right)^{8/3}\frac{1+\Delta}{\Delta}\tau^{-5/3}}{\left(\frac{3\pi}{4}\right)^{8/9}\Delta^{-1/3}\tau^{-8/9}\left(\dot{y}-\frac{8}{9}\frac{y(\tau)}{\tau}\right)},\nonumber
\end{equation}
we have
\begin{equation}\label{eq:dgp_ss_bc_rho1}
    D_1(\lambda_s)  = - \frac{2}{9} \tau_s \frac{1+\Delta} {y_s^2\left(\dot{y}_s - \frac{8}{9}\frac{y_s}{\tau_s}\right)}.
\end{equation}
For $P$, we have $P_1=0$ again.

The following steps are the same as before. We can use the Rankine–Hugoniot jumping conditions to obtain $D_2, P_2, V_2$ and $M_2$, and numerically integrate the equations again to find the postshock solutions. This time we need to vary $\tau_s$ for our trial-and-error process after $\Delta$ is specified. $y_s$ and $\dot{y}_s$ can be calculated numerically from $\tau$ by using Eq.~\eqref{eq::newton_gravity} for the EdS model and Eq.~(\ref{eq:rescaled_DGP_gravity}) for the DGP model. For EdS, we have explicitly checked that using the $\tau$-based approach to set up the boundary conditions for the postshock solutions gives identical answer as using Eqs.~(\ref{eq:pre_shock_soln_D} - \ref{eq:pre_shock_soln_M}), as expected.

We summarise our result for self-similar solutions in DGP gravity with in Figure \ref{fig:self-similar_unrescaled}. 
The black curves in the figure are the self-similar solutions to $D, P, V$ and $M$ for the EdS model, which we find to be in excellent agreement with literature results \citep[e.g.,][]{1985ApJS...58...39B}. The coloured curves show the results for several variants of the `artificial' DGP model described in Section \ref{subsubsect:dgp_spherical}, with the case $\zeta=0$ (red) corresponding to a constant enhancement of $G$ by $1/(3\beta)$. The cases with $\zeta=1$, $5$ and $10$ represent progressively more efficient Vainshtein screening, which explains why they are in between the EdS and $\zeta=0$ cases. In particular, we see that at $\zeta=10$ the screening is already very efficient so that the brown curves are very close to EdS. The qualitative trend also agrees with what one should expect for a model with enhanced gravity: the infall becomes faster such that the preshock solution of $V$ becomes more negative and the shock happens at larger radius; the density $D$ and pressure $P$ are also higher due to the stronger structure formation, and the latter explains why the enclosed mass $M$ within a given radius is larger.

\begin{figure*}
    \centering
    \includegraphics[width=\textwidth]{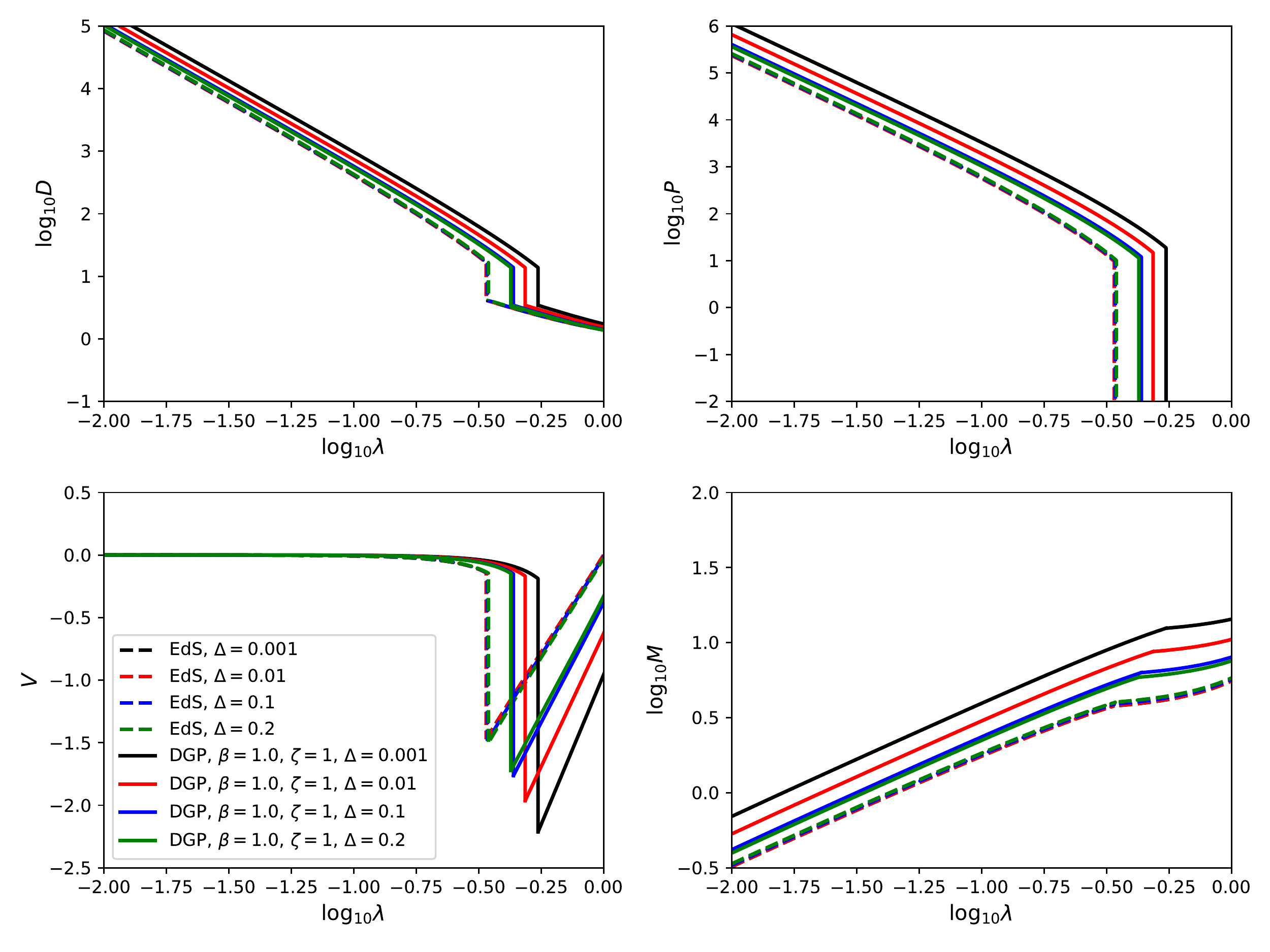}
    \caption{(Colour Online) Two sets of the self-similar solutions are presented here, with the dashed lines for standard gravity in EdS and solid lines for DGP model of $\beta=1.0$, $\zeta=1$. Different choices of $\Delta$ are indicated by colours. The $\Delta$ dependence in EdS case is negligible as the solutions assume the limit $\Delta \ll 1$ \citep{1985ApJS...58...39B}. On the other hand, the results for DGP model shows a clear dependence on $\Delta$, which we explain in the text.} 
    \label{fig:self-similar-unrescaled-Delta-dependence}
\end{figure*}

\begin{figure*}
    \centering
    \includegraphics[width=\textwidth]{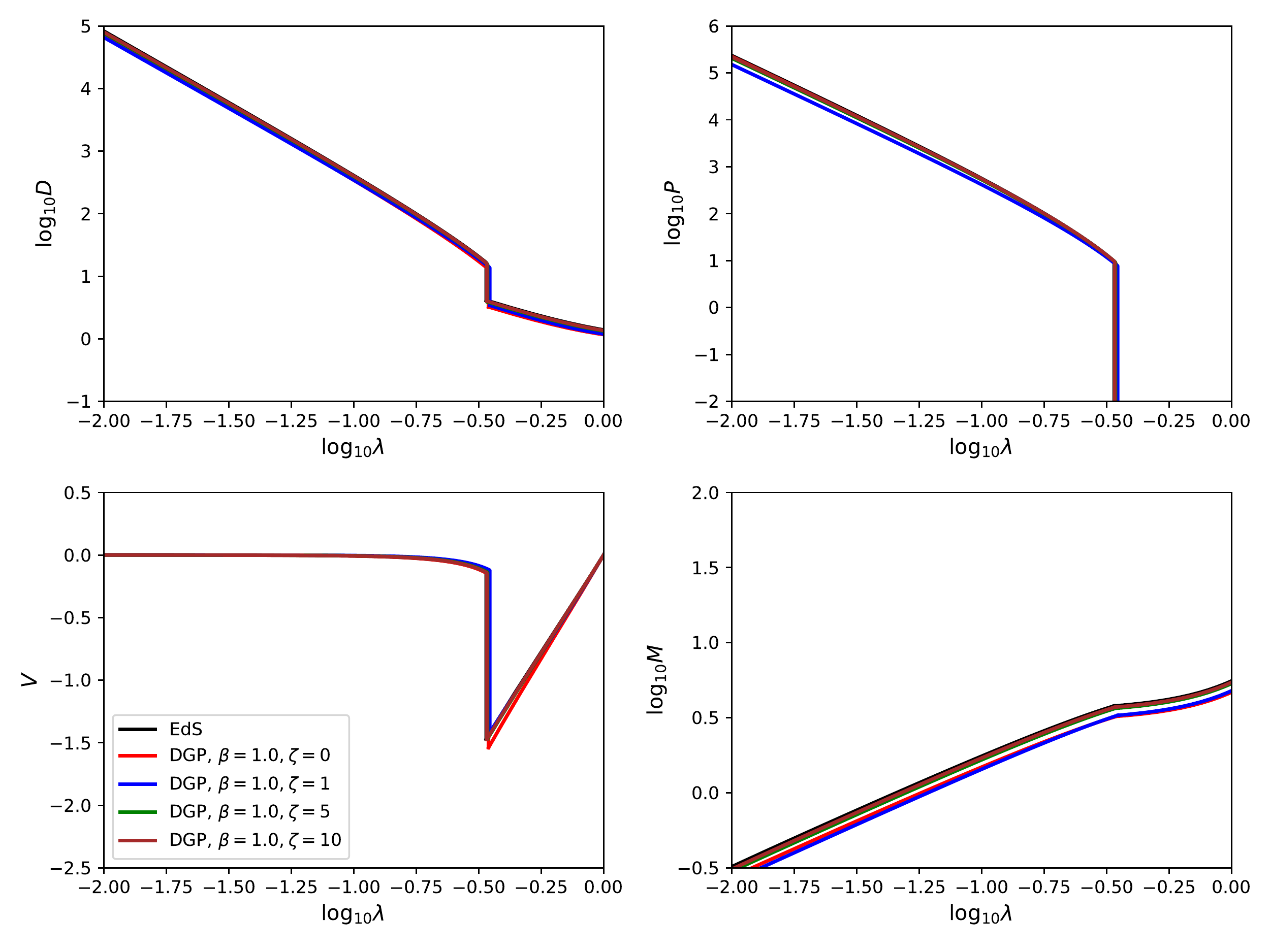}
    \caption{(Colour Online) The new self-similar solutions of DGP models with the same $\zeta$ choices as in previous figures, after the 're-rescaling'. Those curves now has little dependence on $\Delta$, and behave closely to the EdS cases.}
    \label{fig:self-similar_rescaled}
\end{figure*}

The DGP results in Figure \ref{fig:self-similar_unrescaled} are obtained with the parameter $\Delta=0.001$. The $\Delta$ dependence of the solution, which we have explicitly checked, can already be seen at the equation level, cf.~Eqs.~(\ref{eq:dgp_ss_bc_v1}-\ref{eq:dgp_ss_bc_rho1}), and also in 
Fig.~\ref{fig:self-similar-unrescaled-Delta-dependence}, where we show the two sets of self-similar solutions for two models, EdS (dashed lines) and DGP with $\beta=1.0, \zeta=1$ (solid lines). Different colours indicates the value of $\Delta$ for each curve, as shown by the legends. We can see that the DGP model has a very strong $\Delta$ dependence. A similar dependence is also present in the EdS case, but is much weaker there --- indeed, it is known that in EdS there is approximately no $\Delta$ dependence in the limit $\Delta\ll1$ \citep{1985ApJS...58...39B}. This $\Delta$ dependence comes from our choice
of using the turnaround radius $r_{\rm ta}(\tau)$, given in Eq.~(\ref{eq::turn_around_radius}), to define the dimensionless coordinate $\lambda$, where $r_{\rm ta}$ itself depends on $\Delta$. For the EdS model, rescaling $r$ using this turnaround radius helps to cancel out the $\Delta$ dependence from Eq.~(\ref{eq:rescaled_newton_gravity}), because this $r_{\rm ta}$ is calculated from the same dynamical equation and has the physical meaning of where the shell start to collapse. But such a cancellation should not be expected to happen when we use the same Eq.~\eqref{eq::turn_around_radius} to define $\lambda$ for the DGP (and generally other gravity) models, since it does not represent the true turnaround radius anymore. 

Using the same $r_{\rm ta}$ to define $\lambda$ in all models above certainly has its advantages. One of these is that Eq.~\eqref{eq::turn_around_radius} is an analytical function with a power-law dependence on $\tau$, which is convenient when deriving the dimensionless equations governing the self-similar evolution. It also allows these equations to take the similar form between the DGP and EdS models. For example, Eq.~(\ref{eq:eq1_dimensionless}--\ref{eq:eq4_dimensionless}) remain almost the same for the DGP model, with only some slight changes of Eq.~\eqref{eq:eq2_dimensionless} to Eq.~\eqref{eq:eq2_dimensionless_dgp2}. In addition, Figure \ref{fig:self-similar_unrescaled} clearly shows the effect of modified gravity law on the collapse of collisional gas and on the formation of shock: this also benefits from the fact that we have used the \textit{same} `turnaround' radius, $r_{\rm ta}(\tau)$, to define the rescaled quantities in \textit{all} models, so that the differences in the rescaled quantities reflect the differences in the same quantities pre-rescaling. Nevertheless, for theoretical interest, we also want to see the results when we actually define $\lambda$ using the \textit{true} turnaround radius of each model. Because there are no analytical expressions for $r_{\rm ta}$ for the DGP model, this has to be done in a ``post-processing'' way: after getting the profile $V(\lambda)$ by following the above steps, we can obtain the real turnaround radius in the preshock $V(\lambda)$ solution, by looking for the value of $\lambda'_{\rm ta}$ where $V(\lambda'_{\rm ta})$ crosses 0; we then get the correct turnaround radius as:
\begin{equation}
    r'_{\rm ta}=\left(\frac{\lambda'_{\rm ta}}{\lambda_{\rm ta}}\right)r_{\rm ta} 
    \equiv \alpha r_{\rm ta},
\end{equation}
and use $r'_{\rm ta}$ to rescale our solutions for the other quantities, which is equivalent to performing the following `re-rescaling':
\begin{eqnarray}\label{eq:ss_solns_rerescaling}
    \lambda &\rightarrow& \alpha^{-1}\lambda,\nonumber\\
    D &\rightarrow& D,\nonumber\\
    P &\rightarrow& \alpha^{-2}P,\nonumber\\
    V &\rightarrow& \alpha^{-1}V,\nonumber\\
    M &\rightarrow& \alpha^{-3}M.
\end{eqnarray}
The new result is summarised in Figure \ref{fig:self-similar_rescaled}. While we only show the results obtained by using $\Delta=0.001$ here, we find that using $\Delta=0.01$, $0.1$, $0.2$ give very similar results. 
One notable property is that the new rescaled profiles are very close to that in Einstein-de Sitter universe, i.e., the DGP model behaves similarly to 
standard gravity if expressed in terms of the $\lambda$ coordinate which is defined using the true turnaround radius of the model. As the real physical evolutions of these models are very different, this similarity is quite interesting, since it suggests that self-similarity works (at least to a good approximation) in more general models than just EdS.

As we shall see below, this ``re-rescaling'' idea using the true turnaround radius can also be applied to the numerical simulation result from \exahype{} 2, and help to check its reliability on handling this scenario.

%% file: exahype.tex
\section{Numerical simulations with \exahype{} 2}
\label{sec3}
In this section, we first introduce the numerical code we implement on \exahype{} 2, then describe how we configure the spherical collapse scenario with it.

Our simulations are based upon an adaptive Cartesian mesh hosting a
Finite Volume discretisation with an explicit Euler.
The code is realised through 
\exahype, which is a publicly available engine designed for generic hyperbolic PDEs that arise in different branches of sciences and engineering. 
We rely on the second-generation \exahype{} 2 code which is a rewrite
that has been used for astrophysical challenges before \citep[e.g.,][]{Reinarz:2019:ExaHyPE}.

\subsection{Spatial and temporal discretisation}
\label{subsect:num_algorithm}

\exahype{} 2 constructs the spatial discretisation from a spacetree formalism
\citep{Weinzierl:19:Peano} combined with block-structured adaptive mesh refinement
\citep[AMR;][]{Dubey:16:SAMR}:
The computational domain is embedded in a cube and split into three equal parts along each coordinate axis. This yields $3^3=27$ smaller cubes. We continue recursively, i.e., decide for each cube whether to cut it into 27 subcubes again. The refinement decision or criterion is subject of discussion below. The process yields an adaptive refined Cartesian mesh. Starting from an initial adaptive mesh, dynamic adaptivity could be realised by applying the splitting in between time steps to yield a finer mesh.

Each cube hosts a $p \times p \times p$ Cartesian mesh. We call these Cartesian meshes patches and make them carry the actual solution representation: each mesh element in the patch holds a piecewise constant solution of the governing equations, i.e., defines one ``finite volume''. Every patch thus consist of $p^3$ volumes. The patch of volumes is augmented with a ``halo\footnote{Note that the word `halo' here is a technical term indicating an extra layer of volumes surrounding each patch, and differs from its usual meaning in cosmology, e.g., dark matter haloes.} layer'' of width one around it. The patches hence yield a non-overlapping domain decomposition of the computational domain, while the haloes introduce an overlap between them.

Let the vector $\vec{Q}: \mathbb{R}^{d} \times \mathbb{R}^{+} \mapsto \mathbb{R}^5$ denote the unknowns of interest as they evolve over time, where the symbol $\vec{~}$ highlights that this is a data (rather than space) vector that in our case has a dimensionality of 5. We approximate the time derivatives with forward finite differences, i.e., $\frac{{\rm d}\vec{Q}}{{\rm d}t} \approx \frac{\vec{Q}^{\text{new}}-\vec{Q}^{\text{old}}}{\delta T}$ with a given time step size $\delta T$ and $\vec{Q}^{\rm old}$, $\vec{Q}^{\rm new}$ representing the values of $\vec{Q}$ at the start and end of the time step. Our equations are a set of generic first-order hyperbolic PDEs
\begin{equation}\label{eq:PDE_original_formulation}
    \frac{{\rm d} \vec{Q}}{{\rm d} t}+\boldsymbol{\nabla}\cdot\boldsymbol{F}(\vec{Q})
    =S(\vec{Q}), 
\end{equation}
where $\boldsymbol{F}(\vec{Q})$ and $S(\vec{Q})$ are the flux and source term, respectively. Here we have used bold symbols to denote space vectors to distinguish them from the notation for data vectors introduced above. 
The generic first-order hyperbolic PDEs can be written in a weak formulation for one timestep as
\begin{eqnarray}
  \int_{\Omega\times[T,T+\delta T]}\frac{{\rm d} \vec{Q}}{{\rm d} t}\chi{\rm
  d}\boldsymbol{x}{\rm d}t &=&
  -\int_{\Omega\times[T,T+\delta
  T]}\boldsymbol{{\nabla}\cdot\boldsymbol{F}}(\vec{Q})\chi{\rm d}\boldsymbol{x}{\rm d}t 
  \nonumber\\
  && +  
  \int_{\Omega\times[T,T+\delta T]}S(\vec{Q})\chi{\rm d}\boldsymbol{x}{\rm d}t,
\label{eq:PDE_weak_formulation}
\end{eqnarray}
where ${\rm d}\boldsymbol{x}$ runs over the domain $\Omega$, $[T, T+\delta{T}]$ denotes the time interval and $\chi(\boldsymbol{x},t)$ is a test function. Eq.~(\ref{eq:PDE_weak_formulation}) needs to hold for arbitrary $\chi$ to fulfil Eq.~(\ref{eq:PDE_original_formulation}).

In \exahype{} 2, we adopt the Rusanov Finite Volume solver \citep{leveque_2002} to solve the Riemann problem that arises once we assume that the solution remains constant within every timestep and every volume $v$, and set all test function as characteristic function of one finite volume, i.e., they are $\chi_v(\boldsymbol{x},t)=1$ within $v$ and vanish anywhere else. The integration of Eq.~(\ref{eq:PDE_weak_formulation}) over time gives us

\begin{eqnarray}\label{eq:PDE_derivation1}
  &&\frac{1}{\delta T}\int_{v}{\rm d}\boldsymbol{x}\left[\vec{Q}(T+\delta T)  - \vec{Q}(T)\right]\nonumber\\ 
   &=& \int_{v}{\rm d}\boldsymbol{x}\left[-\boldsymbol{\nabla}\cdot\boldsymbol{F}\left(\vec{Q}(T)\right) + S\left(\vec{Q}(T)\right) \right]\nonumber\\
   &=& 
   \int_{v}S\left(\vec{Q}(T)\right){\rm d}\boldsymbol{x} - \oint_{\partial{v}}\boldsymbol{F}(T)\cdot{\rm
   d}\boldsymbol{S},
\end{eqnarray}
where ${\rm d}\boldsymbol{S}$ is the (oriented) area element of the surface of the volume $v$, $\partial{v}$. 
Here the closed-surface integration is decomposed into the summation of multiple faces that have constant normal vector respectively. 
At the same time, we assume the solution vector $\vec{Q}$ to be piece-wise constant, so we apply the following replacement:
\begin{equation}
    \int_v \bigcirc{\rm d}\boldsymbol{x} \to \bigcirc  V_v,\quad 
     \oint_{\partial v}
    \boldsymbol{\bigcirc} \cdot {\rm d}\boldsymbol{S}\to
    \sum_{\partial v} \boldsymbol{\bigcirc}\cdot \boldsymbol{n}S_{\partial v},
\end{equation}
where $V_v$ is the volume of $v$, $S_{\partial v}, \boldsymbol{n}$ are the
area and unit normal vector of one face of $\partial v$, respectively, and $\bigcirc$ ($\boldsymbol{\bigcirc}$) denotes a generic scalar (space vector) function. This leads to final explicit Euler time stepping scheme we implemented in the code:
\begin{eqnarray}\label{eq:solver_time_stepping}
    \forall_v:~\vec{Q}(T+\delta T)-\vec{Q}(T)=S(\vec{Q})T V_v+\sum_{\partial v} {\rm Flux}^{\pm}(\vec{Q})\Big| _{\partial v} T S_{\partial v},
\end{eqnarray}
with the so-called Rusanov flux \citep{rusanov61}:
\begin{eqnarray}\label{eq:rusanov_flux}
    {\rm Flux}^\pm(\vec{Q})\Big|_{\partial v} &=&\frac{1}{2}\left(F_{\boldsymbol{n}}(\vec{Q}^+)+F_{\boldsymbol{n}}(\vec{Q}^-)\right)\nonumber\\
    && -{\rm max}\left(\lambda_{\rm max}(\vec{Q}^+),\lambda_{\rm max}(\vec{Q}^-)\right)\left(\vec{Q}^+-\vec{Q}^-\right).
\end{eqnarray}
$F_{\boldsymbol{n}}$ is the flux term evaluated along ${\boldsymbol{n}}$ of the considered $\partial v$ for the respective volume. 
${\rm Flux}^\pm(\vec{Q})$ in Eq.~(\ref{eq:rusanov_flux}) averages component-wisely over the flux within the two adjacent volumes. 
The average then is corrected (limited):
$\lambda_{\rm max}$ is the largest eigenvalue of the matrix $A(\vec{Q})$ acting on the gradient along $\boldsymbol{n}$ if we write down the PDE along the face normal as
\begin{equation}\label{eq:PDE_gradient_formulation}
    \frac{{\rm d} \vec{Q}}{{\rm d} t}
    \Big|_{\boldsymbol{n}}
    + A(\vec{Q}) \frac{{\rm d} \vec{Q}}{{\rm d} x_{\boldsymbol{n}}} = ...\textbf{}
\end{equation}

\noindent
It indicates the largest propagating speed of the quantities in the system. 

We close this subsection by briefly commenting that $\delta T$ is subject to the Courant–Friedrichs–Lewy (CFL) condition with 
\begin{equation}
 \label{equation:exahype:CFL-condition}
 \delta T < C \frac{|\delta v|}{\lambda_{\rm max}},
\end{equation}
where $C<1$ is a problem-specified safety parameter. Our scheme employs a global time stepping scheme and thus uses the smallest
global face length $|\delta v|$. It remains invariant over time as we fix the finest resolution in our simulations. 
The maximum eigenvalue $\lambda_{\rm max}$, however,  changes over time and thus has to be recalculated after each step.

\subsection{Implementation}

\begin{figure}
    \centering
    \includegraphics[width=200px]{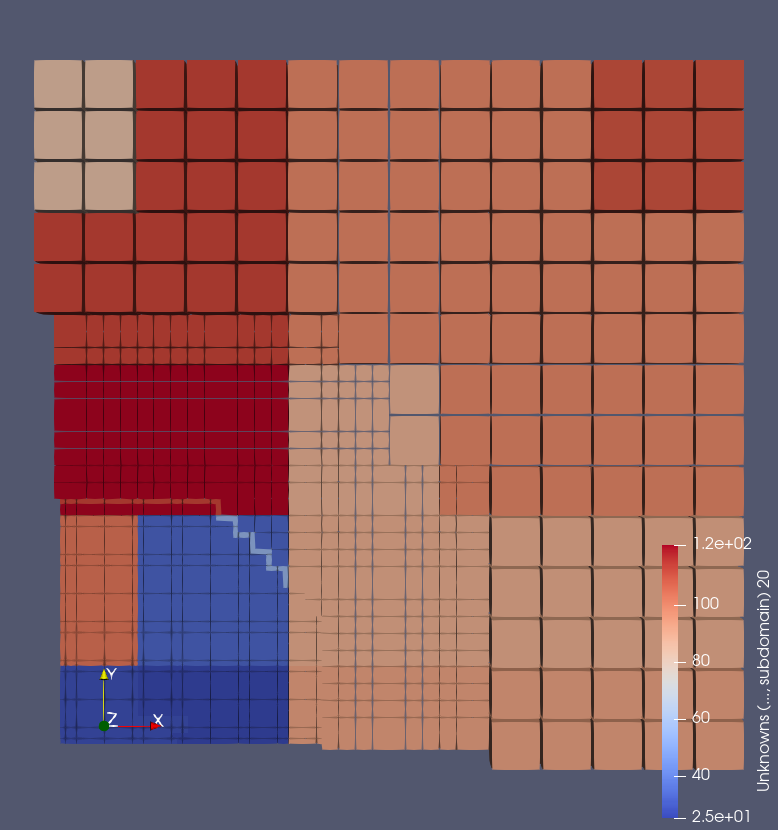}
    \caption{
      (colour online) A two-dimensional cut through one quadrant of the computational mesh. The colours represent subdomains handled by different threads of different ranks.
      As the illustration is a cut-through, the space-filling curve structure is not visible directly.
    }
    \label{fig:SFC}
\end{figure}

Our code splits up the computational domain along the Peano space-filling curve (SFC) into subdomains \citep{Li:22:ISC,Weinzierl:19:Peano} (Fig.~\ref{fig:SFC}): 
All patches are ordered along the SFC.
We cut this sequence of patches into segments such that each rank gets exactly one segment hosting roughly the same number of patches.
As the Peano SFC is continuous, the set of patches per rank form a connected subdomain of the computational domain which does not overlap with any subdomain handled on another rank.
Per rank, we apply the SFC splitting once more such that each thread per rank obtains its
own subdomain:
The patches within the computational domain are first distributed among the ranks and each rank then distributes its patches once more among the threads.
This gives us a two-level non-overlapping MPI+OpenMP parallelisation.

Our realisation with patches supplemented with a halo of width one allows each thread to run through the mesh, and to update all of its patches independently of the other ones. After this mesh traversal, the halo layers are copied over for neighbouring patches of the same size, halo finite volumes overlapping with coarser resolution patches are updated due to a linear interpolation, while halo volumes
overlapping with finer resolutions are updated through averaging over the finer volumes. We map the individual patch updates per thread onto a task formalism
\citep{Li:22:ISC} and process the patches along the MPI boundaries prior to other tasks such that the data transfer required for the halo updates can overlap with further computations \citep{Charrier:2020:EnclaveTasking}. 
We assume that the
tasks can compensate for any geometric ill-balancing on the MPI level. We do not dynamically rebalance throughout the computation.

In our experiments, we use four nodes of Durham's COSMA 7 cluster with one MPI rank per compute node.
Each node hosts a dual-socket Intel Gold 5120 CPU processor.
Therefore, each rank splits up its domain into 28 further subdomains. 
Our experiments stick to $p=3$.
While this setup yields a relatively low arithmetic load per patch compared to the overhead that we need to maintain the halo volumes, it ensures that we can use a rather aggressive coarsening towards the domain boundaries to reduce the overall computational burden.


\subsection{Code Units}
\label{label:code_units}

To solve the system of equations numerically, it is usually convenient to recast them by using dimensionless quantities. In the \exahype{} 2 implementation, we adopt the so-called supercomoving coordinates, which are used in other simulation codes such as \ramses{} \citep{ramses_paper}.

The original formulation of this coordinate system could be found in \cite{Martel98}. Its idea is to apply the following rescaling of the variables:
\begin{eqnarray}\label{eq:supercomoving_unit}
    {\rm d}\tilde{t} &\equiv& H_0 \frac{{\rm d}t}{a^2},\nonumber\\
    \tilde{x} &\equiv& \frac{1}{a}\frac{x}{L},\nonumber\\
    \tilde{\rho} &\equiv& a^3 \frac{\rho(\boldsymbol{x},t)}
    { \Omega_{{\rm m0}}\rho_c}~~=~~\frac{\rho(\boldsymbol{x},t)}{ \bar{\rho}_{\rm m}(t)},\nonumber\\
    \tilde{p} &\equiv& a^5 \frac{p}{\Omega_{\rm m0}\rho_c H^2_0 L^2},\nonumber\\
    \tilde{\boldsymbol{u}} &=& a\frac{\boldsymbol{u}}{H_0L}.
\end{eqnarray}
Here $\rho_c, \bar{\rho}_{\rm m}(t)$ are respectively the critical density today and mean density of matter at 
time $t$; $L$ is the comoving size of unit code length; ${\rm d}t$, $x$ and $\boldsymbol{u}$ denote, respectively, the (physical) time interval, physical coordinate and peculiar velocity. We use the quantities with a tilde in our code, we therefore call them code unit in the following context. 

The supercomoving coordinate system factors out most of the effect from the Hubble expansion, and thus allows us to implement the original fluid equations Eq.~(\ref{eq::gas_eq1}-\ref{eq::gas_eq4}) in a static space with just minor changes. For the special case $\gamma=5/3$, the only change of the fluid equations is a re-calibration of the gravity term in Eq. (\ref{eq::gas_eq2}), which now needs to be derived from the following code-unit 
Poisson equation:
\begin{equation}\label{eq:poisson_code_unit}
    \tilde{\nabla}^2 \tilde{\Phi}=\frac{3}{2}\Omega_{\rm m0} a(\tilde{\rho}-1),
\end{equation}
where $\tilde{\Phi}$ is the Newtonian potential in code unit
\begin{equation}
    \tilde{\Phi} = \frac{a^2\Phi}{L^2H_0^2}.
\end{equation}
Solving Eq.~\eqref{eq:poisson_code_unit} 
under spherical symmetry gives us the following solution of the Newtonian gravitational force $\tilde{g}\equiv-{\rm d}\tilde{\Phi}/{\rm d}\tilde{r}$ (again, in code unit):
\begin{equation}\label{eq:gravity_term_code}
    \tilde{g}=-\frac{3}{2}\Omega_{\rm m0} a\frac{1}{\tilde{r}^2}\int^{\tilde{r}}_0\big[\tilde{\rho}(\tilde{r}')-1\big]\tilde{r}'^2 d\tilde{r}'\equiv -\frac{3}{2}\Omega_{\rm m0}a\frac{1}{\tilde{r}^2}\frac{\delta \tilde{M}(<\tilde{r})}{ 4\pi},
\end{equation}
where we have defined 
$\delta \tilde{M}(<\tilde{r})$ to be the total ``mass perturbation'' within radius $\tilde{r}$, i.e., the difference between the total mass therein and the mass in the same region were the density there equal to $\bar{\rho}_{\rm m}$. For other fluid equations, we only need to replace physical quantities with code quantities directly. For cases $\gamma \neq 5/3$, extra terms are needed for supercomoving coordinates (although they are straightforward to derive), which we do not cover here. 

The generalisation to calculate the modified gravitational force in the DGP model is straightforward: we multiply the fifth-force-to-Newtonian-gravity ratio $\xi$ given in Eq.~(\ref{eq:DGP_gravity_modifier}) to Eq.~(\ref{eq:gravity_term_code}) directly to obtain the 
fifth force in the DGP model. Most terms in Eq.~\eqref{eq:DGP_gravity_modifier} are constants or time-dependent functions, and the only term that needs to be rewritten in code unit is 
\begin{equation}\label{eq:xi_term_code1}
    \frac{1+\Delta}{y^3} \tau^2 -1 
    = \frac{\frac{4\pi}{3}\rho_ir_i^3\left(1+\Delta\right)}{\frac{4\pi}{3}\rho_ir^3}\tau^2-1 = \frac{m(r_i,t_i)}{\frac{4\pi}{3}\rho_ir^3}\tau^2-1,
\end{equation}
where we recall that $r_i$ is the initial radius of the 
fluid element located at $r$ at time $t$, and $m\left(r_i,t_i\right)$ is the total mass enclosed within $r_i$ at the initial time $t_i$. As no shell crossing happens during the evolution, the mass within the radius of this fluid element 
remains the same, which means:
\begin{equation}\label{eq:xi_term_code2}
     \frac{1+\Delta}{y^3} \tau^2 -1 = \frac{m(<r,t)}{\frac{4\pi}{3}\rho_ir^3}\tau^2-1,
\end{equation}
where $m(<r,t)$ denotes the total mass enclosed in radius $r$ at time $t>t_i$. In our code implementation, the mass is calculated by counting 
volumes (see Section \ref{subsect:sim_settings} below), and thus
$m(<r,t) = \sum_{r_k\leq r}\rho_k(t)\ell_k^3$, where the subscript $k$ labels the volumes, $\ell_k$ is the cubic size of volumes $k$, and $\rho_k(t)$ is the density (all in physical units). Notice that we have:
\begin{equation}\label{eq:xi_term_density_relation}
    \frac{\rho_k(t)}{\rho_i}\tau^2 = \frac{\tilde{\rho}_k \rho_{\rm H}(t)}{\rho_i \tau^{-2}} = \frac{\tilde{\rho}_k \rho_{\rm H}(t)}{\rho_{\rm H}(t)}=\tilde{\rho}_k
\end{equation}
in the Einstein-de Sitter universe. Putting Eq.~(\ref{eq:xi_term_density_relation}) back to Eq.~(\ref{eq:xi_term_code2}), we get
\begin{equation}\label{eq:xi_term_code_final}
    \frac{1+\Delta}{y^3} \tau^2 -1 =\frac{\sum_{r_k<r} \tilde{\rho}_k \ell_k^3}{\frac{4\pi}{3} r^3}-1
    =\frac{\sum_{r_k<r} (\tilde{\rho}_k-1) \ell_k^3}{\frac{4\pi}{3} r^3}
    =\frac{\delta \tilde{M}(<\tilde{r})}{\frac{4\pi}{3} \tilde{r}^3},
\end{equation}
where in the second equality we have used $\frac{4\pi}{3}r^3=\sum_{r_k\leq r}\ell_k^3$, while in the final equality we have replaced $\ell_k$ and $r$ with their code-unit expressions, $\tilde{\ell}_k$ and $\tilde{r}$, which does not change the ratio $\ell^3_k/r^3$, and used $\delta\tilde{M}(<\tilde{r})\equiv \sum_{\tilde{r}_k<\tilde{r}}\left(\tilde{\rho}_k-1\right)\tilde{\ell}_k^3$. Eq.~(\ref{eq:xi_term_code_final}) is the final code expression that we use in our simulation.

\subsection{Simulation Settings}
\label{subsect:sim_settings}

In this subsection, we discuss how we implement the spherical collapse scenario on \exahype{} 2. We describe the hyperbolic equations and grid setting that are used in the simulations, the initial conditions and boundary conditions, and how we calculate the total perturbed mass, $\delta \tilde{M}(<\tilde{r})$, at arbitrary radius $\tilde{r}$.

\subsubsection{Equations and Grid setting}
\begin{figure*}
    \centering
    \includegraphics[width=\textwidth]{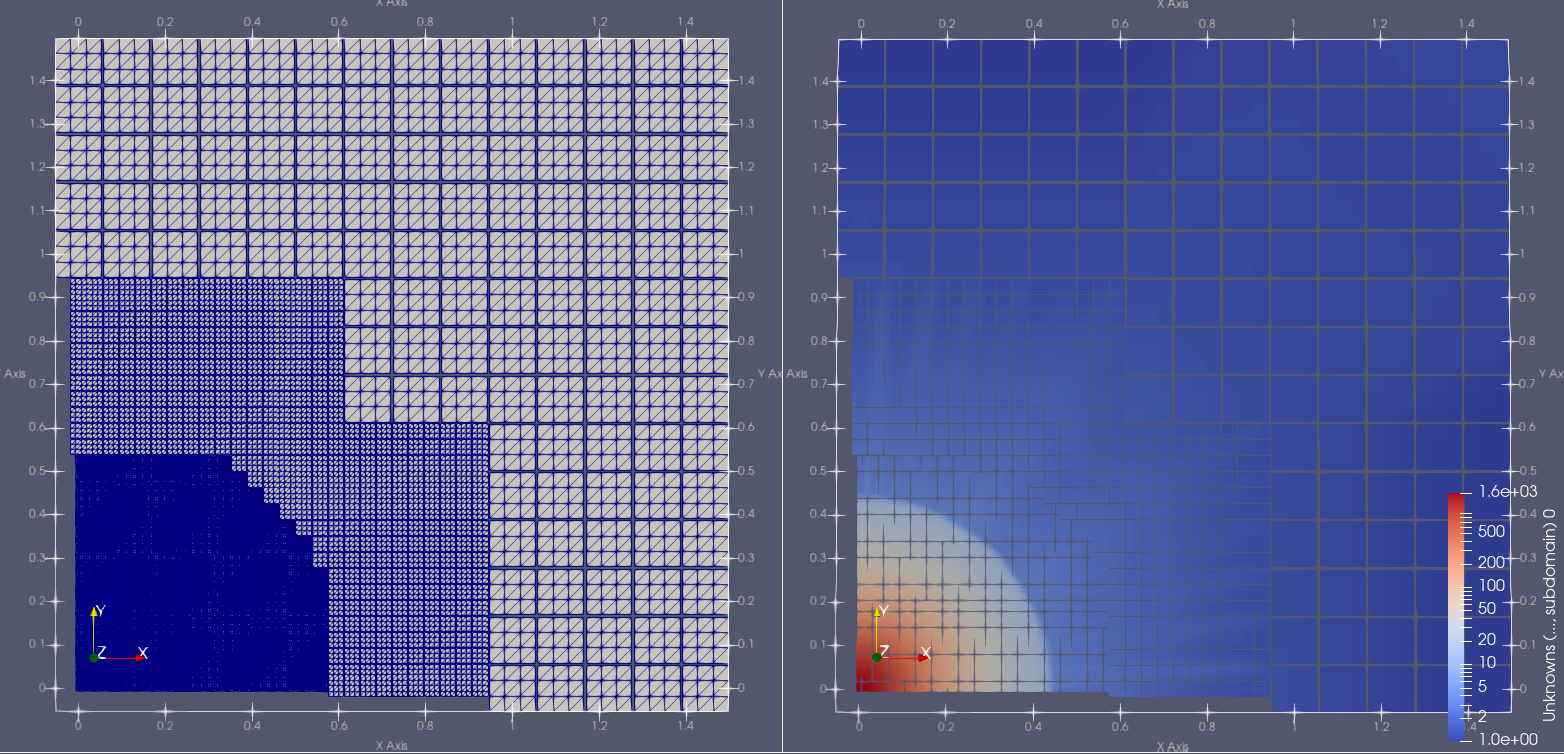}
    \caption{(Colour Online) \textit{Left Panel:} The adaptive Cartesian grid used in our simulations, with patches and volumes that we describe in Section \ref{subsect:num_algorithm} therein. The patches with $p=3$ (i.e., every patch contains $3^3$ volumes) are separated from each other in the visualisation with gaps for clarity. Three levels of the grid are shown here. Only one quarter of $x$-$y$ plane
    taken from a slice of the simulation box perpendicular to the $z$-axis is plotted. The diagonal lines are visualisation artefacts as we use the cubic finite volumes. The refinement transitions are conservative, i.e., they are slightly larger than the resolution transitions imposed by the refinement strategy.
    \textit{Right Panel:} The density field (in code unit) on the same slice for a snapshot during a simulation. Some fluctuations of the density field could be seen out of the central peak, as we discuss them in Section \ref{subsubsec:bc}.
    }
    \label{fig:amr_setting}
\end{figure*}
In the simulations, we implemented the original conservation form of the (gravity-driven) Euler equations in code unit:
\begin{eqnarray}
\frac{\partial \tilde{\rho}}{\partial\tilde{t}}+\nabla\cdot\tilde{\boldsymbol{j}} &=& 0,
\label{eq:euler_eq_rho}
\\
 \frac{\partial \tilde{\boldsymbol{j}}}{ \partial\tilde{t}} + \nabla\cdot \left(\frac{1}{\tilde{\rho}}\tilde{\boldsymbol{j}}\otimes\tilde{\boldsymbol{j}}+\tilde{p}\mathbf{I}\right) &=& \tilde{\boldsymbol{f}},\label{eq:euler_eq_j}
 \\
  \frac{\partial \tilde{E}}{ \partial \tilde{t}} +\nabla\cdot\left(\frac{1}{\tilde{\rho}}\tilde{\boldsymbol{j}}(\tilde{E}+\tilde{p})\right) &=& \frac{1}{\tilde{\rho}}\tilde{\boldsymbol{j}}\cdot\tilde{\boldsymbol{f}},
 \label{eq:euler_eq_energy}
\end{eqnarray}
where $\tilde{\rho}$, $\tilde{\boldsymbol{j}}$, $\tilde{E}$, $\tilde{p}$ represent the density of mass, momentum, energy and pressure in code unit respectively, $\tilde{\boldsymbol{f}}=\tilde{\rho}\tilde{\boldsymbol{g}}$ is the force density with $\tilde{\boldsymbol{g}}$ the gravitational acceleration, which is proportional to $\delta \tilde{M}(<\tilde{r})/ \tilde{r}^2$.
We consequently obtain $\vec{Q}=(\tilde{\rho},\tilde{\boldsymbol{j}},\tilde{\boldsymbol{E}})$
in Eq.~(\ref{eq:PDE_original_formulation}).

All simulations we presented in this paper use the same grid setup on a cubic
box $[-1.5,1.5]^3$. The maximum
refinement level within the tree formalism is 3, corresponding to a
resolution of $243^3$ patches on the finest level. Every patches contains 27 volumes again ($p=3$). We coarsen this mesh once at a distance of $0.5$ (in code units) away from the origin, and coarsen it once more at $0.7$. Figure \ref{fig:amr_setting} illustrates the AMR refinement pattern we used for the simulation. The exact refinement pattern is chosen such that it covers the refinement radii.
The safety parameter (CFL ratio) we use in Eq.~(\ref{equation:exahype:CFL-condition}) is $C=0.3$.

\subsubsection{Initial Conditions}

The simulations shown in this paper start at scale factor $a_i=0.001$, and end around $a \approx 0.3$. 
The simulation domain is initially filled with collisional cold gas of $\gamma=5/3$ in critical density (which is unity in code units). Our overdense seed, the spherical tophat, is placed at the origin and is set to have a radius $\tilde{R}_i=0.05$ and total 
perturbed mass $\delta \tilde{M}_i=0.15$.

The treatment of the initial conditions of the pressure, density and velocity is subtle.
Although we should expect a pressureless infall for most regions in the simulation box at the beginning, we can not set a zero initial pressure numerically.
Likewise, although it seems to be quite natural to set a zero initial velocity profile within our comoving coordinate system, we can not do this in our implementation, neither. Both of these would lead to a negative pressure in the first time step. This is because in this step the energy equation, Eq.~(\ref{eq:euler_eq_energy}), does not update the local energy given the zero momentum (i.e., both the flux and the source terms are zero in this equation). On the other hand, the momentum itself is updated normally according to Eq.~(\ref{eq:euler_eq_j}) as its source term (the force density) is nonzero. Since we calculate the pressure using:
\begin{equation}
    \tilde{p}=(\gamma-1)\left(\tilde{E}-\frac{1}{2}\tilde{\mathbf{j}}^2/\tilde{\rho} \right),
\end{equation}
the fact that $\tilde{\boldsymbol{j}}$ is updated (mostly increased in magnitude) while $\tilde{E}$ is not during the first step can cause an accidental and unphysical drop of pressure at the end of this timestep, and frequently (for zero initial pressure, it is always) the pressure turns to be negative where gravity is strong, i.e., near the centre. This issue would be worse if we put a point mass as the overdense seed at the centre, like the one in \ramses{} \citep{ramses_paper}, because it leads to an extremely large magnitude of the gravity force in the adjacent volumes of the point mass. 

To address this negative pressure issue, our solution is three-fold. Firstly, we stick to using a tophat overdensity rather than a point mass as our seed, though it harms the solution partially (see the section for results below). A tophat initial profile smooths the gravity field and reduces the magnitude of a potential negative pressure. Secondly, we set a very small but non-zero value for the pressure initially: it makes the system more robust to the pressure drop in the first time step, and can quickly converge to the correct pressureless solution outside the shock later in the simulation. Finally, we introduce a pre-set initial velocity profile. We assume our momentum field has evolved a small period of (physical) time before the simulation begins, according to the initial gravity field:
\begin{equation}\label{eq:pre_set_velocity}
    \tilde{\boldsymbol{j}}_i=\Delta_t \tilde{\boldsymbol{g}}_i,
\end{equation}
such that the energy can get updated as well. These adaptions successfully solve the initial negative pressure issue without the explicit construction of consistent initial condition which does not yield unphysical solutions. The freedom of adjusting our initial conditions without harming the final self-similarity is  
expected given the convergence of the solution \citep{alard20}, and we have explicitly checked that it is true for our simulation by tuning our initial pressure.

\subsubsection{Boundary Conditions}
\label{subsubsec:bc}

\begin{figure}
    \centering
    \includegraphics[width=150px]{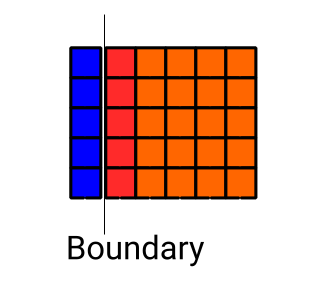}
    \caption{(Colour Online) An illustration of the boundary layout used in our simulations with \exahype{} 2. Outside the boundary of the simulation domain, denoted by the thin black line, a layer of ghost volumes (blue squares) are set up, and the interested quantities in the ghost volumes, $\vec{Q}_{\rm out}$, depend only on the values of these quantities, as well as their first derivatives, in the volumes immediately inside the boundary (dark red squares), $\vec{Q}_{\rm in}$. See Eqs.~(\ref{eq:boundary_condition_1}, \ref{eq:boundary_condition_2}) for the exact details.}
    \label{fig:ghost_cell}
\end{figure}

Our setup to simulate spherical collapses requires free inflow boundary conditions. Because we expect $\vec{Q}$ to be almost stationary in comoving coordinates (or approaching the Hubble flow physically) as we move away from the centre of the computational domain, homogeneous Neumann boundary conditions can yield the free inflow as long as the computational domain is 
sufficiently large. However, such a large domain is computationally inefficient or even unfeasible, and it is also not clear whether `large' is well-defined in an evolving system: the shock propagates outwards towards the border over time, thus making it a challenge to use homogeneous Neumann boundary conditions throughout the entire evolution. We therefore use the following hybrid scheme:
\begin{numcases}
{\vec{Q}_{\rm out}=}
    \vec{Q}_{\rm in}, & $\tilde{\rho}_{\rm in}<1 $
    \label{eq:boundary_condition_1}
    \\
    \vec{Q}_{\rm in}+\tilde{\ell}_{\rm in} ~
    \boldsymbol{n}\cdot{\displaystyle\lim_{\boldsymbol{x}\to \partial \Omega}}\boldsymbol{\nabla}^{(1)}\vec{Q}(\boldsymbol{x}), & {\rm otherwise},
    \label{eq:boundary_condition_2}
\end{numcases}
where $\vec{Q}_{\rm in}$ and $\vec{Q}_{\rm out}$ denote, respectively, the solution vectors in the volumes on the inner and outer sides of the boundary (see Fig.~\ref{fig:ghost_cell}). The boundary conditions in \exahype{} 2 are implemented by specifying how the quantities in ghost volumes out of the boundary $\vec{Q}_{\rm out}$ are calculated from ones in their direct neighbours within the domain $\vec{Q}_{\rm in}$. In most times, we use the extrapolating boundary condition Eq.~(\ref{eq:boundary_condition_2}), where 
the superscript $^{(1)}$ means we use the first-order approximation of the gradient $\boldsymbol{\nabla}\vec{Q}$ at $\boldsymbol{x}$ approaching the domain boundary $\partial\Omega$, multiplied by the distance between the two volumes, $\tilde{\ell}_{\rm in}$. 

\begin{figure*}
    \centering
    \includegraphics[width=\textwidth]{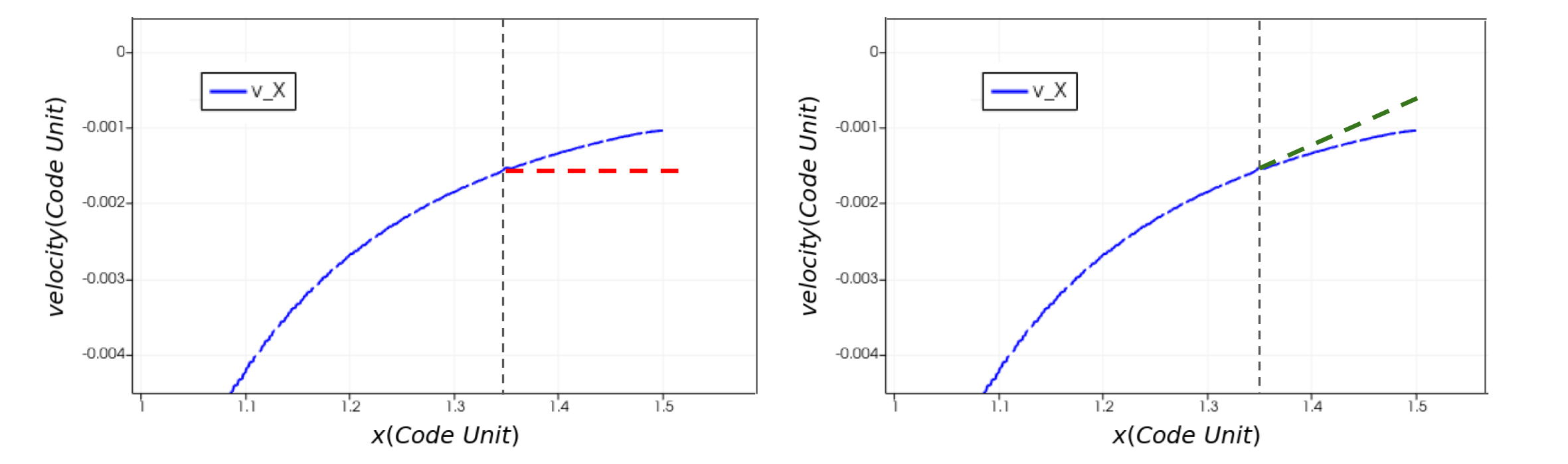}
    \caption{(Colour Online) The different 
    ways to set up the boundary conditions in the velocity field at the boundary of the simulation domain, indicated by the vertical dashed line. In both panels, the blue curve illustrates a physical velocity profile along the $x$ direction, which has a nonzero gradient at the boundary. \textit{Left panel:} the homogeneous Neumann boundary condition specified by Eq.~(\ref{eq:boundary_condition_1}), where the velocity field outside the boundary (i.e., to the right of the vertical line) is assumed to be a constant equal to the velocity value just inside the boundary (the red dashed line). In this case, the inflow from beyond the boundary is overestimated, and thus harms the quality of the boundary. \textit{Right panel:} The first-order extrapolated boundary condition corresponding to Eq.~(\ref{eq:boundary_condition_2}), as indicated by the green dashed line. Its prediction of inflow is more accurate than the Neumann case but is underestimated. We combine these two boundary conditions in our simulations depending on the local density at the boundary.}
    \label{fig:bc_hybrid}
\end{figure*}

The different behaviours of these two types of boundary conditions are illustrated in Fig.~\ref{fig:bc_hybrid}. The linearly extrapolated boundary condition is 
more accurate than the homogeneous Neumann one specified by Eq.~(\ref{eq:boundary_condition_1}), but it underestimates the momentum inflow from beyond the boundary. As a result, the code-unit density at the boundary, $\tilde{\rho}_{\rm in}$, will drop to under unity later in the evolution: this is unphysical because the density everywhere in this collapse scenario should be above the critical density. Whenever this happens, we switch to the homogeneous Neumann boundary condition, Eq.~(\ref{eq:boundary_condition_1}). The 
latter usually overestimates the inflow, and thus can 
provide some `compensation'. After the density $\tilde{\rho}_{\rm in}$ increases back to above unity, we continue using the extrapolating boundary condition again.


\vspace{0.3cm}

The Finite Volume scheme uses normal boundary conditions where the normal is axis-aligned. However, our solution is spherical-symmetric.
The boundary condition's normal alignment thus is erroneous. Even with Eqs.~(\ref{eq:boundary_condition_1}--\ref{eq:boundary_condition_2}),
we have to ensure that the domain remains sufficiently large compared to the area of interest, such that this misalignment becomes negligible. This naturally limits the maximum simulation time up to which our results are not polluted significantly by the tangential boundary errors, as the solution's steep gradient moves towards the domain boundary.

Similar arguments hold along resolution transitions. As we interpolate linearly along the resolution boundary, our solutions do not follow exact spherical symmetry: the mesh and its resolution transitions should be spherical, and we should interpolate linearly along a spherical transition.
Yet, our grid is Cartesian.
This ``misalignment'' results in fluctuations or finger patterns (Fig.~\ref{fig:amr_setting}, right panel). Our code has two ingredients to mitigate the resulting error: on the one hand, we use 2:1 balancing \citep{Sundar:08:BalancedOctrees}, since a more aggressive resolution change would amplify any error. On the other hand, we ensure that the ``first'' (finest to second finest) resolution transition is
sufficiently far away from the region of interest, i.e. the shock. In return, this implies that the maximum runtime yielding physically admissible results is bounded further, as long as we disable adaptive mesh refinement---a technique which is intrinsically limited, as the area of interest expands and thus eventually yields a regular grid with excessive memory footprint.

\subsubsection{Mass Integration}

Most of the terms in Eqs.~(\ref{eq:euler_eq_rho}-\ref{eq:euler_eq_energy}) can be implemented in \exahype{} 2 directly as part of the Rusanov scheme on Cartesian meshes we describe above, because they are all localised variables, i.e. follow up the update pattern of any Finite Volume scheme. However, the gravitational force
\begin{equation}
    |\boldsymbol{\tilde{f}}| = \tilde{\rho}|\tilde{\boldsymbol{g}}| = \frac{3}{2}\Omega_m\tilde{\rho}a\frac{\delta \tilde{M}(<\tilde{r})}{4\pi\tilde{r}^2}(1+\xi),
\end{equation}
is not localised as we will need the total perturbed mass within radius $\tilde{r}$. To get 
$\delta \tilde{M}(<\tilde{r})$, we construct a mass array $\{\delta\tilde{m}_i\}_{0 \leq i \leq i_{\rm max}}$ which stores the total perturbed mass values within radii $\{\tilde{r}_i\}_{0 \leq i \leq i_{\rm max}}$. Here $\tilde{r}_{\text{max}}=\tilde{r}_{i=i_{\rm max}}$ is chosen to be the radius of the largest sphere in the simulation box: half of the domain length. The values of $\delta\tilde{m}_i$ are calculated by accumulating the mass in all volumes that are within $\tilde{r}_i$ per time step:
\begin{equation}
    \delta\tilde{m}_i(<\tilde{r}_i,t) = \sum_{\tilde{r}_k\leq \tilde{r}_i} \left[\tilde{\rho}_k(t)-1\right] \tilde{\ell}_k^3,
\end{equation}
where $\tilde{\ell}_k$ is the size of the accumulating volume. The plain summation is consistent with our choice of piece-wise constant Finite Volumes. During the subsequent time step, we apply the following interpolation rule per volume according to its radius $\tilde{r}$ for the required perturbed mass:
\begin{equation}
    \delta \tilde{M}(<\tilde{r})=\begin{cases}
    \delta\tilde{ m}_0 \tilde{r}^3/ \tilde{r}^3_0,& \tilde{r} \leq \tilde{r}_0\\
    \delta\tilde{ m}_{i}\left(\frac{\tilde{r}_{i+1}-\tilde{r}}{ \tilde{r}_{i+1}-\tilde{r}_i}\right)+\delta\tilde{m}_{i+1}\left(\frac{\tilde{r}-\tilde{r}_i}{ \tilde{r}_{i+1}-\tilde{r}_i}\right),& \tilde{r}_{i}<r\leq \tilde{r}_{i+1}\\
    \delta\tilde{m}_{\text{max}}+\frac{4\pi}{3} \tilde{\rho}(\tilde{r}_{\text{max}})\left(\tilde{r}^3-\tilde{r}^3_{\text{max}}\right). & \tilde{r}>\tilde{r}_{\text{max}}
    \end{cases}
    \label{eq:mass_interpolation}
\end{equation}
The perturbed masses for volumes outside 
$\tilde{r}_{\rm max}$ are approximated by assuming that the density there is equal to that at $\tilde{r}_{\rm max}$. During our simulations, 
the densities in those volumes depart little from unity and thus contribute little to the total perturbed mass. This approximation 
is therefore acceptable. More accurate schemes could be used in future simulations, such as using a scheme of density interpolation that can extend to the furthest corner of the simulation box. Within  
$\tilde{r}_{\rm max}$, on the other hand, the accuracy of this interpolation 
rule depends on the size and arrangement of the sample array $\{\delta\tilde{m}_i\}$. In our simulations, we use a sample array size of 200, and keep our sample radii $\{\tilde{r}_i\}$ invariant over time.

%% file: result_and_conclusion.tex
\section{Simulation Results}
\label{sec4}

In this section, we report the simulation results of spherical collapse scenarios in different gravity models using our new code. To make 
comparison to the theoretical predictions we got in Section \ref{sec2}, 
we will also show results that are rescaled following Eqs.~(\ref{eq:dimensionless_v}-\ref{eq:dimensionless_m}), after we restored the quantities in physical unit using Eq.~(\ref{eq:supercomoving_unit}). 

\subsection{Einstein-de Sitter universe}

\begin{figure*}
    \centering
    \includegraphics[width=\textwidth]{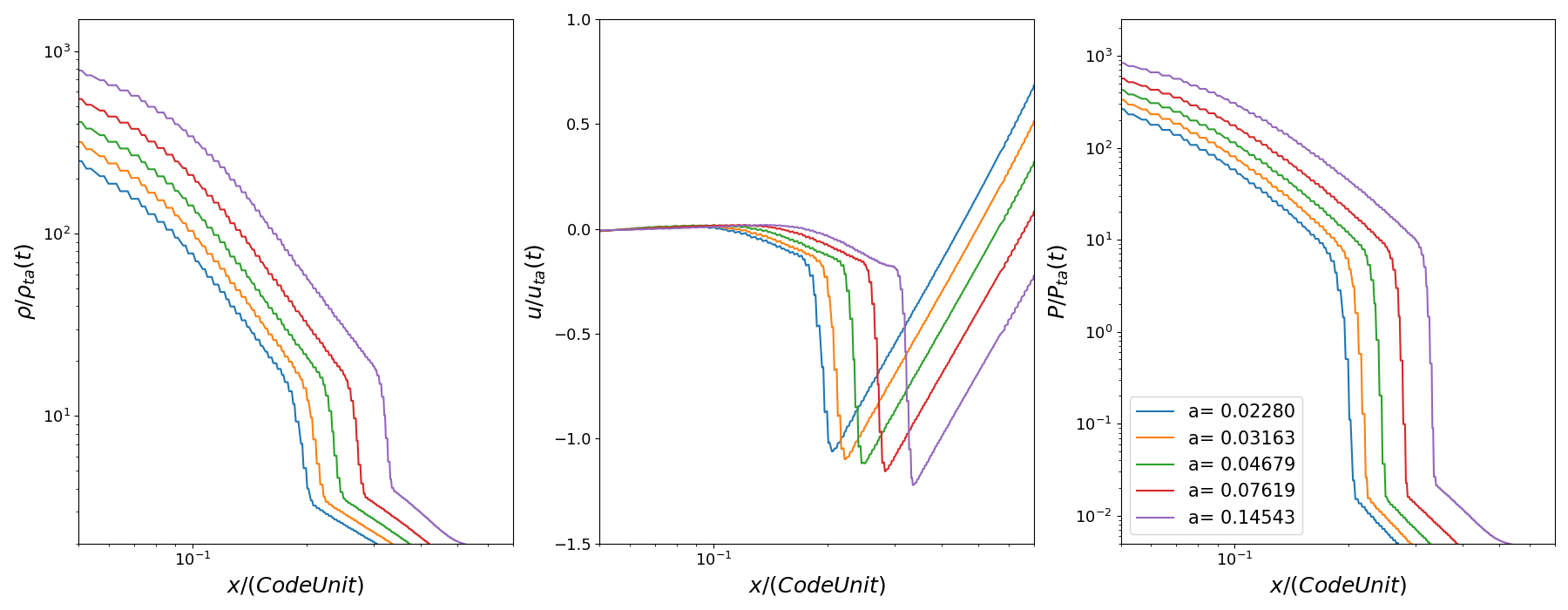}
    \caption{(Colour Online) The rescaled density, velocity and pressure profiles for spherical collapse in an Einstein-de Sitter universe, 
    plotted against the radius from the centre in code units. 
    Five snapshots of the system at $a\approx 0.022$, $0.031$, $0.047$, $0.076$, $0.145$ are shown in different colours as indicated by the legends. An outward-propagating shock is clearly visible in all three panels. The curves are sampled over the positive direction of the $x$ axis, but we have checked that for all the simulations we report in this section
    the solution only has a very weak dependence on the direction along which we extract it from the simulation domain, see Appendix \ref{app:spherical_symmetry}. 
    }
    \label{fig:eds_codeX}
\end{figure*}

We first show the simulation results in the Einstein-de Sitter universe. 
Since gravity is standard, we can use Eq.~(\ref{eq::turn_around_radius})
as the scaling radius. The rescaled profiles of physics quantities are plotted over the radius coordinates in the code unit (supercomoving coordinates), in Fig.~\ref{fig:eds_codeX}. We illustrate five snapshots of the system (at scale factor $a\approx 0.022$, $0.031$, $0.047$, $0.076$, $0.145$) 
from the late part of the simulation when the corresponding $\Delta$ is relatively small. The system remains in stable evolution before the numerical issues we reported in the last section pollute the solution. A clear outward-propagating shock can be seen in the figure.

\begin{figure*}
    \centering
    \includegraphics[width=\textwidth]{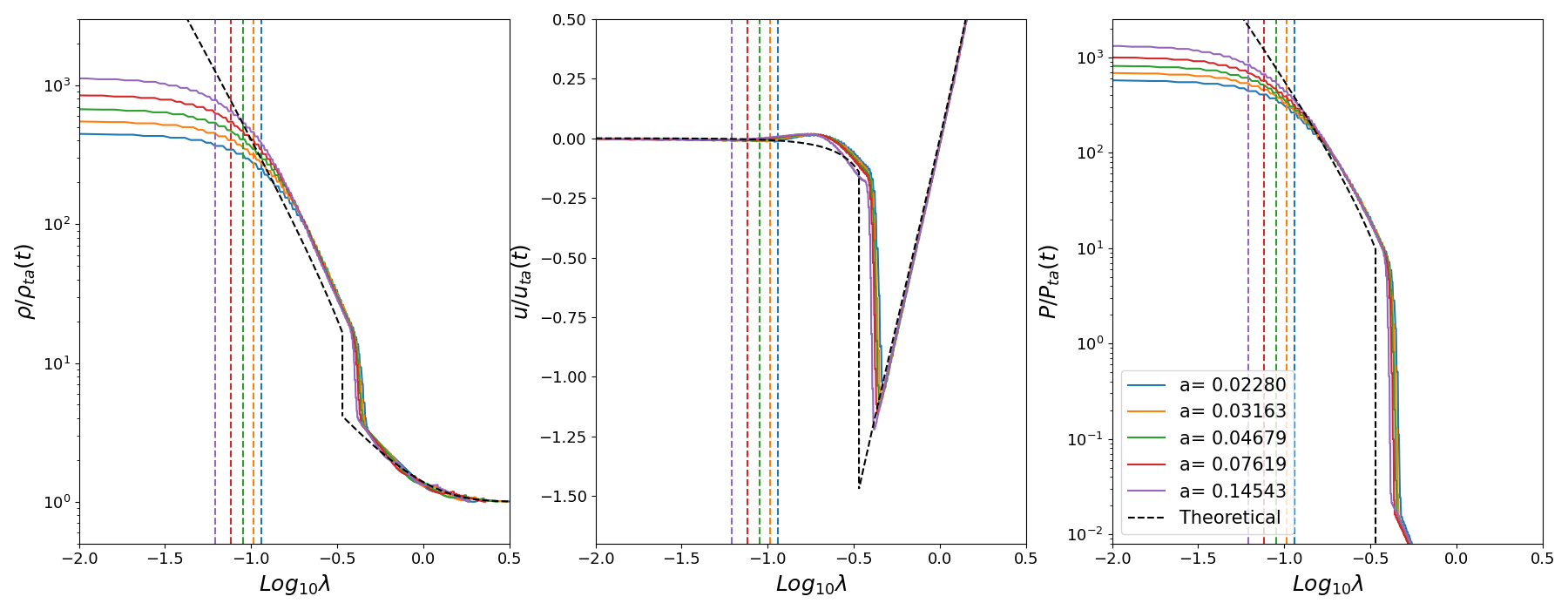}
    \caption{(Colour Online) The rescaled density, velocity and pressure profiles from the the same simulation of the Einstein-de Sitter model, 
    plotted against the rescaled radial coordinate, $\lambda$. The self-similar theoretical prediction \citep{1985ApJS...58...39B} is 
    shown as black dashed lines. The vertical dashed lines with colours 
    indicate the locations of the tophat edge at the same five times as shown in Fig.~\ref{fig:eds_codeX}, 
    and the numerical solutions depart from the self-similar 
    prediction within it. This location is moving inwards as the rescaling radius $r_{\rm ta}$ increases over time.
    Convergence over time to the theoretical solution can be observed in the plots.}
    \label{fig:eds_rescaledX}
\end{figure*}

The same profiles of quantities are plotted again, but now 
against the rescaled radial coordinates $\lambda$, in Fig.~\ref{fig:eds_rescaledX}. The theoretical self-similar lines from Section \ref{sec2} are shown as black dashed lines for comparison. We can see a clear self-similarity here, as the rescaled simulated quantities 
have converged during the time period considered, when the scale factor $a$ increases by a factor of seven.
The coloured vertical lines in the figures are the positions of the 
tophat edge at the time of the corresponding snapshots, within which the density and pressure solution deviate from the self-similar solution and flatten: this is expected as the gas within the tophat does not experience the full gravity from the mass perturbation anymore. The radius of this edge is shrinking in the rescaled plots over time because the turnaround radius that is used to define $\lambda$ increases as time evolves. 

The rescaled solutions agree with the theoretical predictions quite well, especially for the preshock solutions of the density and velocity. Yet, there are some deviations from the self-similar solution, notably a shift of the shock position. Because of this, the infall velocity of the gas just outside the shock is lower than the theoretical prediction. This is a common numerical artefact caused by volumes with finite widths, which cannot exactly resolve the infinitesimally thin shock. We have checked that the agreement with the self-similar prediction improves as we use finer volumes. A detailed convergence study is beyond scope here.

Another factor that may have contributed to the
difference between theory and simulation is that the theoretical solution
here is obtained under the assumption of $\Delta \ll 1$, and this is not
well satisfied in the simulations. 
The different shells of gas have different initial radii $r_i$ and corresponding values of $\Delta$, with the outer shells having larger $r_i$ and therefore smaller $\Delta$, and vice versa. The outer shells also collapse to the shock at a later time. We note that the outer shells that collapse to the shock at later stages of the simulations 
usually have $\Delta \approx 0.1$, and the inner shells have even larger $\Delta$. The difference between these values of $\Delta\ll1$ might affect the accuracy of the simulation results.  
This claim is supported by the time convergence of the
profiles in Figure \ref{fig:eds_rescaledX} toward the self-similar solutions. However, it is not clear to what extent letting the simulation run for longer, so that shells with ever larger $r_i$ will fall to the shock, helps here, since some of the inaccuracy of the simulation results is due to numerical dissipation. Additionally, as we explained in Section \ref{subsubsec:bc}, the maximum runtime yielding physically admissible results is bounded, and simulations after a longer time will begin to depart from the self-similar solution generally. 

\begin{figure*}
    \centering
    \includegraphics[width=\textwidth]{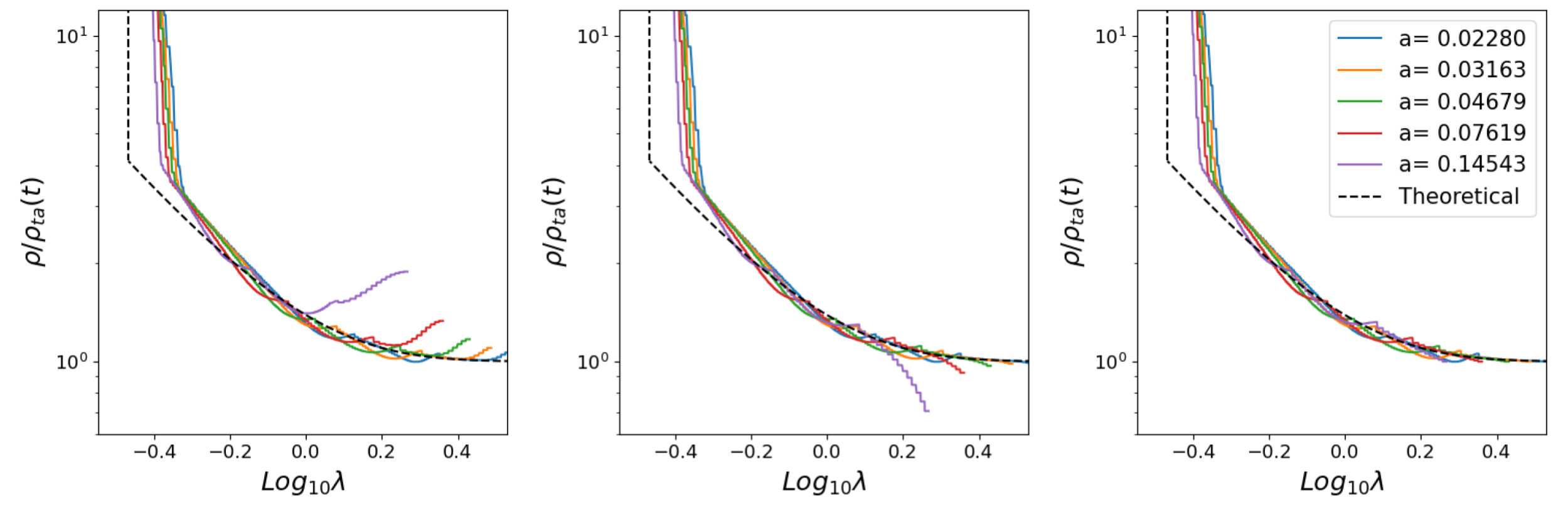}
    \caption{(Colour Online) The tail parts of the rescaled density 
    profiles from three EdS 
    simulations which implement different boundary conditions but are otherwise identical. The three panels, from left to right, show the results from using homogeneous Neumann (outflow), pure extrapolated and hybrid boundary conditions, respectively. Only in the hybrid case does the profile near the boundary remain stable and consistent with the theoretical prediction (the dashed line). The other two cases either overestimates or underestimates the density near the boundary, leading to an eventual crash of the simulation when the error near the boundary propagates into the central region of the simulation domain (see the purple and red solid lines in the left two panels).}
    \label{fig:eds_bc}
\end{figure*}

We next study the effect of hybrid boundary condition scheme introduced in Section \ref{subsubsec:bc}. Figure \ref{fig:eds_bc} gives the tail part of the density profiles of three simulations which are identical except for the implementations of the boundary conditions. The three 
panels correspond to the three types of boundary conditions mentioned above, respectively homogeneous Neumann (outflow), pure linear extrapolation and the hybrid scheme. 
A clear abnormal uprising of density near the boundary 
can be seen in the homogeneous case (the first panel), as it overestimates the inflow from beyond the boundary. 
This effect would ``propagate'' inwards and 
eventually pollute the solution, making it unstable. On the other hand, the density drops to under unity (or the critical density in physical units) when we use the extrapolated boundary condition (the second panel), 
leading to a negative density later in the simulation. By using 
the hybrid scheme (the third panel), we manage to keep a relatively stable and smooth density evolution near the boundary throughout the simulation. 

\subsection{DGP models}

In this subsection we report the simulation results of the DGP model introduced in Sections \ref{subsect:dgp_model} and \ref{subsubsect:dgp_spherical}, with $\beta=1.0$ and various values of the screening parameter $\zeta$. For a 
clear comparison with the standard gravity, we first rescale our modified gravity results using the same turnaround radius formulation Eq.~(\ref{eq::turn_around_radius}), following what we did first in Section \ref{subsect:ss_solns}. As the rescaling radius is identical in the different gravity models, the differences after the rescaling also represent the difference in the real evolution, thus showing the effects of modified gravity and the screening mechanism.

\begin{figure*}
    \centering
    \includegraphics[width=\textwidth]{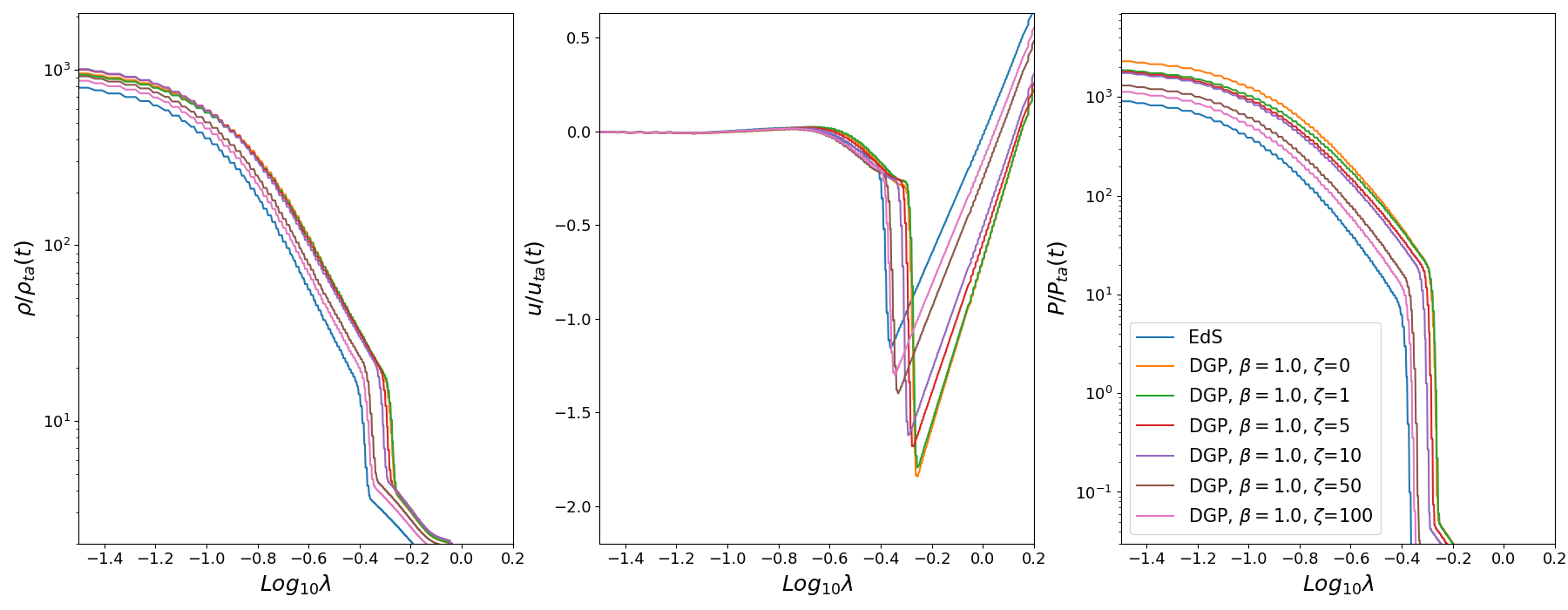}
    \caption{(Colour Online) The rescaled quantities curves in the DGP model introduced in Sections \ref{subsect:dgp_model} and \ref{subsubsect:dgp_spherical}, with $\beta=1.0$ and $\zeta=0$, $1$, $5$, $10$, $50$, $100$, at the time step when $a\approx 0.076$. The profiles in an EdS universe are also illustrated here for a comparison (blue solid lines). The case of $\zeta=0$ corresponds to a model with a constant (in space and time) enhancement of Newton's constant, while $\zeta>0$ introduces the Vainshtein screening effect which grows with $\zeta$. It is 
    therefore as expected that the case with a constant enhancement of gravity ($\zeta=0$) deviates most from the EdS result, and results of other cases lie in between.
    }
    \label{fig:dgp_unrerescaled}
\end{figure*}

The results at $a\approx 0.076$ for models with $\beta=1.0$ and $\zeta=0$, $1$, $5$, $10$, $50$, $100$ are summarised in Figure \ref{fig:dgp_unrerescaled}. 
Those results agree with what one should expect for an enhanced gravity force and presence of screening: 
for the non-screening case ($\zeta=0$), in which 
gravity is constantly enhanced in time and space, a 
stronger shock is observed and it also happens at a larger radius than in EdS. In the 
other cases, as the screening becomes stronger and stronger (i.e., 
increasing $\zeta$), the results approach that of standard gravity in an EdS universe. 

\begin{figure*}
    \centering
    \includegraphics[width=\textwidth]{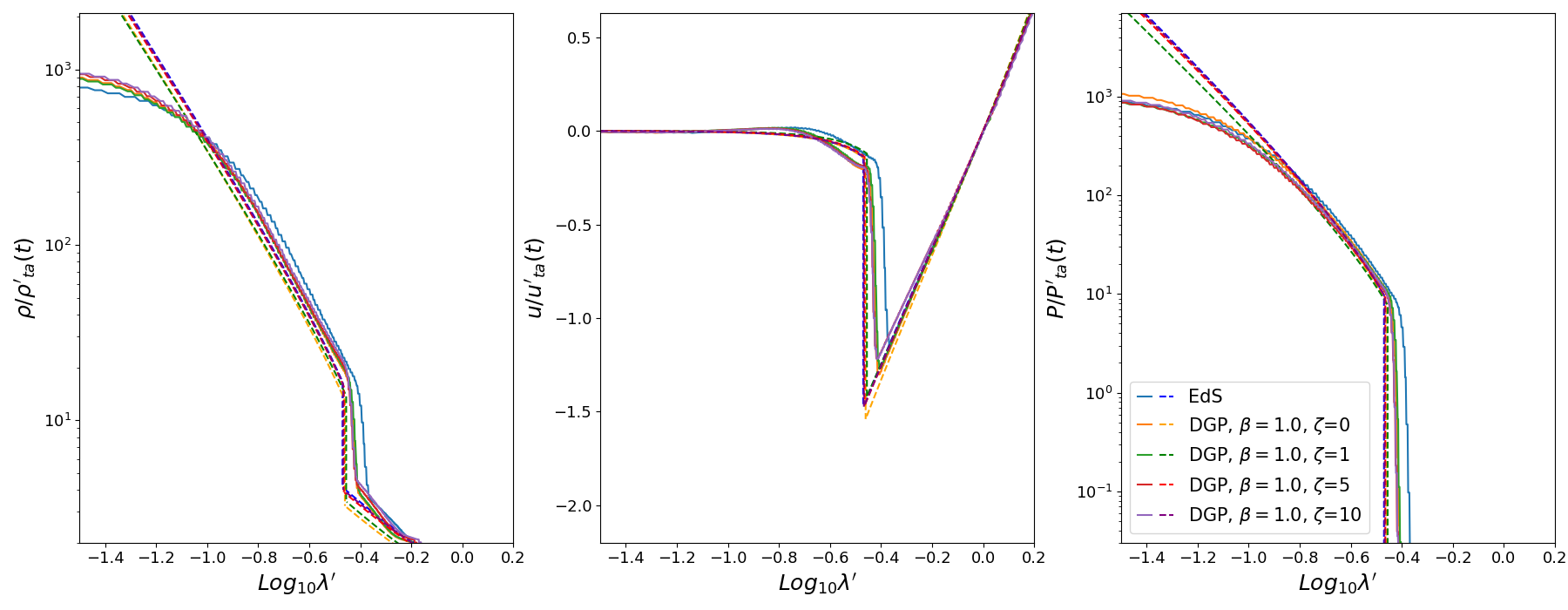}
    \caption{(Colour Online) The profiles of the same quantities 
    as shown in Fig.~\ref{fig:dgp_unrerescaled} (solid lines) for the DGP model with different parameters (see legends), now rescaled using the real turnaround radius $r'_{\rm ta}$ as described in Section \ref{subsect:ss_solns}. The quantities 
    with a prime 
    are calculated using this new rescaling radius. We also plot the theoretical self-similar predictions 
    described in Section \ref{subsect:ss_solns} for each case, 
    as dashed lines with the corresponding colours.}
    \label{fig:dgp_rerescaled}
\end{figure*}

One may have noticed that we require a bigger $\zeta$ to achieve a similar screening effect, compared to Figure \ref{fig:self-similar_unrescaled}. This is mainly due to the fact that the 
parameter $\Delta$, which characterises the mean initial overdensity density within some given initial radius $r_i$, takes different values at the different initial radii covered by a real simulation, while 
the theoretical profiles are 
obtained assuming a fixed $\Delta$, e.g., $\Delta=0.001$. To get rid of the $\Delta$ dependence in our results, we use the idea of rescaling using the true turnaround radius as described in Section \ref{subsect:ss_solns}. The difference is that this time we do not need a ``re-rescaling'': after we restore the profile quantities in physical units, we find the real turnaround radius directly by its physical meaning, i.e., we locate the radius where the physical velocity crosses zero. This method can be applied to all 
models including the EdS, which we have checked explicitly to give the same result as in the subsection above. After we located this real turnaround radius for simulations with DGP model, we use this value for our rescaling. The result of the same simulations and same timestamp in this new rescaling scheme then are plotted as the solid lines in Fig.~\ref{fig:dgp_rerescaled}, and their theoretical predictions (as shown in Fig.~\ref{fig:self-similar_rescaled}) are overplotted as the dashed lines with the same colour scheme. In the figure titles, we have used primes to indicate the quantities calculated using the numerically determined turnaround radius, $r'_{\rm ta}$. 

Just like the theoretical results we got in \ref{subsect:ss_solns}, the new rescaled solutions are close to that in EdS universe. They are broadly in line with the theoretical predictions as well. The shock in DGP model happens at a slightly smaller radius, and the velocity in the gas shell just outside the shock has a bigger magnitude. 
This result is possibly caused by the fact that the gravitational force is stronger in the DGP model, so that the collapse is also stronger and faster.
The qualitative trend is also as expected, as the curves for the models with screening are between the ones of EdS and a constant enhancement of Newton's constant ($\zeta=0$). Given that the real physical evolutions of these models in the simulation are quite different (cf.~Fig.~\ref{fig:dgp_unrerescaled}), these results demonstrate the reliability of \exahype{} 2 engine to carry 
out both standard and modified gravity simulations, and support the idea that 
self similarity can be found (at least as a very good approximation) in more general gravity models beyond EdS as well.

\section{Discussion and Conclusion}
\label{sec5}
To summarise, we have studied the spherical collapse
of collisional gas in both an Einstein-de Sitter universe and DGP gravity model in this paper. 
We have derived self-similar solutions, for the first time, for
some special cases of the latter class of models. The existence of self-similar solutions in spherical collapse scenarios is nontrivial: for example, while the EdS model admits a self-similar solution, this is lost if the model includes a cosmological constant. This is even more true for modified gravity models, in which the law of gravity may be modified in complicated time- and spatial-dependent ways. Indeed, we have tried to search for self-similar solutions in several classes of modified gravity theories that feature certain screening mechanisms. Chameleon-type models \citep{Khoury:2003rn,Khoury:2003aq} do not admit self-similar solutions, because the fifth force there is not only scale dependent but also environment dependent. We have not found self-similar solutions for K-mouflage-type models \citep[][]{Babichev:2009ee,Brax:2012jr} either: in this model, the fifth force is given by 
\begin{equation}
    F = \beta_{\rm K}\frac{{\rm d}\varphi}{{\rm d}r},
\end{equation}
where $\beta_{\rm K}$ is a parameter describing the coupling strength of the scalar field $\varphi$ with matter, which is usually taken as a constant or function of time. The radial gradient ${\rm d}\varphi/{\rm d}r$ can be schematically obtained by solving 
\begin{equation}
    K\left(\frac{{\rm d}\varphi}{{\rm d}r}\right) \propto \beta_{\rm K}\frac{m(<r)}{r^2} \propto \beta_{\rm K}\frac{r_{\rm ta}}{t^2}\frac{M(\lambda)}{\lambda^2},
\end{equation}
where $K(\cdot)$ is a nonlinear function, and $r_{\rm ta}$ is again the EdS expression of the turnaround radius, Eq.~\eqref{eq::turn_around_radius}. For the fifth force to also respect self-similarity, it should be possible to express it as
\begin{equation}
    F = \mathcal{F}(\lambda)\frac{r_{\rm ta}}{t^2},
\end{equation}
where $\mathcal{F}$ is a function of $\lambda$ only. This is satisfied if $K(\cdot)$ is a linear function and $\beta_{\rm K}$ is a constant, but this simply corresponds to a model with a constant enhancement of Newton's constant, identical to the DGP variant with $\zeta=0$ considered above. For general $K(\cdot)$, one has to require $\beta_{\rm K}$ to depend on both $r_{\rm ta}$ (and through which also depend on the initial radius $r_i$ and overdensity $\Delta$) and $t$ to satisfy the above condition. Even for the DGP model we considered above, demanding a self-similar solution places some constraints on certain details, in particular the requirement that $r_c$ becomes a time-dependent function which grows at the same rate as the horizon size of the EdS universe. The existence of self-similar solutions in specific models offers us a way to test our numerical code for models other than EdS.

The self-similar solutions we obtained for the modified DGP model behave as one would expect for an enhanced gravity with the Vainshtein screening mechanism at work. For example, we see that the shock happens at a larger radii in the DGP variant with $\zeta=0$, and the infall velocity is larger outside the shock, compared with EdS, as a result of a stronger gravitational collapse. For the other DGP variants where $\zeta>0$, the results generally lie between EdS and $\zeta=0$, indicating a suppressed fifth force, and the suppression effect is larger for larger $\zeta$. 
It is notable that, despite the substantial differences in the evolutions and solutions of the different gravity models considered, after the (more `proper') rescaling using the 
true turnaround radius of individual models, the solutions in the different DGP variants are all very close to that in the EdS model with standard gravity (though their agreement is not perfect). We also notice that, after this proper rescaling, the self-similar solutions in the DGP models depend very weakly on $\Delta$, as also happens in EdS. Apparently, we should test these observations for other types of gravity models too. if they hold there as well, this is an interesting indication that the properly rescaled solutions in different gravity models are close to each other, which in turn implies that self-similarity should hold approximately, even though not exactly, in generic models. We leave a more detailed exploration of this possibility to future work.

Behind our new physical insights is a new implementation of cosmological hydrodynamical simulations of the spherical collapse scenario for different gravity models, based on the publicly-available hyperbolic PDE engine \exahype{} 2.
We have described various technical details in our implementation, including the initial and boundary conditions which must be properly set up in order to get stable and correct evolutions. We find that the numerical 
simulations of the same EdS and DGP models as introduced above yield
good agreements with the theoretical predictions we derived. This thus not only supports our 
findings on the self-similarity in the 
considered models, but also serves as a validation of the reliability and correctness of our \exahype{} 2 implementation. 

By comparing our theoretical predictions to the simulation 
results, we find that, although to a large degree 
the code is capable of handling the collapse scenarios in different gravity models,
there are still some inaccuracies in the current simulation results, in particular at and around the shock. 
The observed shift and weakening of the shock are likely caused by numerical dissipation, which 
may be suppressed by increasing the spatial and temporal resolutions.
There are several possible ways of doing this. First, we are currently using a
simple Finite Volume formalism which employs a generic Riemann solver. This scheme can be further extended to 
higher-order formalisms, e.g., Discontinuous Galerkin methods in combination with Runge-Kutta schemes or ADER-DG \citep{zanotti2015space}. Those schemes are in principle compatible with our scenarios, straightforward to implement, and could work properly to enhance the resolutions, but it remains an open question if these methods are well-suited to capture the steep gradients near the shock. We could also directly increase the resolution of our simulations, but this comes at additional runtime cost. Future work will look at the outsourcing of the individual patches to GPUs. This will provide us with the opportunity to work with significantly finer resolutions and a much higher efficiency. For the temporal side, we need to check if local time stepping or subcycling help to reduce the vulnerability of the current explicit time stepping scheme to numerical dissipation. It may also help to use more accurate Riemann solvers, as the current Rusanov solver only `reacts' to the biggest eigenvalue of the system, cf.~Eq.~(\ref{eq:rusanov_flux}), so that it does not preserve the characteristics of all five 
evolving quantities well if they propagate with different wave speeds.

With a working hydrodynamical simulation code at hand, where new models of gravity can be straightforwardly implemented, a natural next step is to run simulations for more realistic modified gravity models that do not have self-similar solutions, including the original DGP model, the K-mouflage model and the chameleon model. For the latter we may need to either add a multigrid solver for the scalar field, or adopt some approximate solutions such as the thin-shell solution. In a future project, we will compare the collapse of collisional gas in these different models in detail. If the above speculation, namely the spherical solutions rescaled by the true turnaround radii of models are approximately the same in different cosmological models, turns out to be correct, then the differences in the physical solutions of these models can be largely ascribed to the differences in their turnaround radii, which might offer a simple way to model the modified gravity effects. In addition, we plan to add more physical processes, such as radiative cooling \citep[e.g.,][]{abadi00}, in the code, to understand how they interfere with the effects of a modified law of gravity. Altogether, these will hopefully offer us new insights into the behaviour of gas, and hence the galaxy formation process, in modified gravity models.

%% file: acknowledgements.tex
\section*{Acknowledgements}

HZ is supported by a Chinese Scholarship Council
(CSC) PhD Studentship, hosted by Durham University. TW and HS acknowledge the support through Durham's oneAPI Academic Centre of Excellence made by Intel, ExCALIBUR's Phase Ia grant ExaClaw (EP/V00154X/1) and ExCALIBUR's cross-cutting project EX20-9 \textit{Exposing Parallelism: Task Parallelism} (Grant ESA 10 CDEL). BL is supported by an European Research Council Starting Grant (ERC-StG-716532), and the UK Science and Technology Funding Council Consolidated Grants No.~ST/I00162X/1 and ST/P000541/1. 

The Exascale Computing ALgorithms \& Infrastructures Benefiting UK Research (ExCALIBUR) programme is supported by the UKRI Strategic Priorities Fund. The programme is co-delivered by the Met Office on behalf of PSREs and EPSRC on behalf of UKRI partners, NERC, MRC and STFC. \exahype{} 2 is currently maintained and extended as part of the embedded CSE programme of the ARCHER2 UK National Supercomputing Service
(\url{http://www.archer2.ac.uk}) under grant no ARCHER2-eCSE04-2 and 
Durham's oneAPI Academic Centre of Excellence made by Intel.

This work used the DiRAC@Durham facility managed by the Institute for Computational Cosmology on behalf of the STFC DiRAC HPC Facility (\url{www.dirac.ac.uk}). The equipment was funded by BEIS via STFC capital grants ST/K00042X/1, ST/P002293/1, ST/R002371/1 and ST/S002502/1, Durham University and STFC operation grant ST/R000832/1. DiRAC is part of the UK National e-Infrastructure.